\documentclass[onecolumn,preprint,preprintnumbers,nofootinbib,floats]{revtex4}
\usepackage{amsmath,amssymb,color,mathrsfs, graphicx,verbatim,epsfig, wasysym}
\usepackage[hyperfootnotes=false]{hyperref}
\usepackage{slashed}
\allowdisplaybreaks

\def\lsim{\mathrel{\rlap{\lower4pt\hbox{\hskip1pt$\sim$}}
    \raise1pt\hbox{$<$}}}
\def\gsim{\mathrel{\rlap{\lower4pt\hbox{\hskip1pt$\sim$}}
    \raise1pt\hbox{$>$}}}

\newcommand{\gev}{{\rm GeV}}

\newcommand{\be}{\begin{eqnarray}}
\newcommand{\ee}{\end{eqnarray}}

\def\addresses#1#2{\hbox to \hsize{\@tablebox{#1}\hfil\@tablebox{#2}}}
\def\@tablebox#1{\vtop{\hsize=5in \begin{flushleft} #1 \end{flushleft}}}

\def\beq{\begin{equation}}
\def\eeq{\end{equation}}
\def\bit{\begin{itemize}}
\def\eit{\end{itemize}}
\def\beqa{\begin{eqnarray}}
\def\eeqa{\end{eqnarray}}

\def\MadGraph{{\tt MadGraph}}
\def\MadGraph5{{\tt MadGraph5}}

\begin{document}

%\baselineskip 0.6cm

%\begin{titlepage}

%\thispagestyle{empty}

%\begin{flushright}
%PAPER ID STUFF
%\end{flushright}

%\begin{center}

%\vskip 2cm

\title{Boosting Top Partner Searches in Composite Higgs Models}

\preprint{IPPP/13/68}
\preprint{DCPT/13/136}

\begin{abstract}
Fermionic third generation top partners are generic in composite Higgs models. They are likely to decay into third generation quarks and electroweak bosons. We propose a novel cut-and-count-style analysis in which we cross correlate the model-dependent single and model-independent pair production processes for the top partners $X_{5/3}$ and $B$. In the class of composite Higgs models we study, $X_{5/3}$ is very special as it is the lightest exotic fermion. A constraint on the mass of $X_{5/3}$ directly extends to constrains on all top partner masses. By combining jet substructure methods with conventional reconstruction techniques we show that in this kind of final state a smooth interpolation between the boosted and unboosted regime is possible. We find that a reinterpretation of existing searches can improve bounds on the parameter space of composite Higgs models. Further, at 8 TeV a combined search for $X_{5/3}$ and $B$ in the $l+\rm{jets}$ final state can be more sensitive than a search involving same-sign dileptons.
\end{abstract}

\author{Aleksandr Azatov} \email{aleksandr.azatov@roma1.infn.it}
\affiliation{Dipartimento di Fisica, Universit$\grave{a}$ di Roma ``La Sapienza" and INFN Sezione di Roma, I-00185 Roma, Italy}

\author{Matteo Salvarezza} \email{matteo.salvarezza@roma1.infn.it}
\affiliation{Dipartimento di Fisica, Universit$\grave{a}$ di Roma ``La Sapienza" and INFN Sezione di Roma, I-00185 Roma, Italy}

\author{Minho Son} \email{minho.son@roma1.infn.it}
\affiliation{Dipartimento di Fisica, Universit$\grave{a}$ di Roma ``La Sapienza" and INFN Sezione di Roma, I-00185 Roma, Italy}

\author{Michael Spannowsky} \email{michael.spannowsky@durham.ac.uk}
\affiliation{Institute for Particle Physics Phenomenology , Department of Physics, Durham University, United Kingdom}

%\end{center}

\maketitle

%\noindent 
%\end{titlepage}

\setcounter{page}{1}

\section{Introduction}
\label{sec:intro}
%%%%%%%%%%%%%%%%%%%%%%
%  Introduction
%%%%%%%%%%%%%%%%%%%%%%

After the recent discovery of the Higgs boson~\cite{Aad:2012tfa,Chatrchyan:2012ufa}, or Higgs-boson-like resonance, the next most important task at the LHC is to unravel the detailed dynamics of the electroweak symmetry breaking mechanism. In absence of any sign of new physics beyond the discovery of the Higgs-like resonance only one guiding principle for the mass scale of new particles remains: naturalness. Or more specifically, are peculiar cancelations necessary to stabilise the mass of the scalar resonance at the electroweak scale? Two approaches are commonly discussed; either the mass is protected by a symmetry (supersymmetry), or the scalar resonance is a composite bound state. In both scenarios the top partners are expected to be light. Hence, searching for light top partners is crucial to understanding whether the naturalness principle is relevant for electroweak symmetry breaking in any way.

In this work we study top partner searches in the context of composite Higgs scenarios~\cite{Kaplan:1983fs,Kaplan:1983sm,Georgi:1984af,Dugan:1984hq,Contino:2003ve}. While the Higgs boson can be a generic composite bound state, we focus on the case where the Higgs boson is realised as a pseudo Nambu-Goldstone boson (pNGB) of the coset ${\mathcal G}/{\mathcal H}$. ${\mathcal G}$ is an approximate global symmetry and ${\mathcal H}$ is an unbroken subgroup (see~\cite{Contino:2010rs} for a review). In order to take into account quarks and leptons, while avoiding being in conflict with constraints from flavour physics~\cite{KerenZur:2012fr}, we assume partial compositeness~\cite{Kaplan:1991dc}. In this picture the Standard Model (SM) fermions get their masses through mixing with the composite bound states from the strongly coupled sector. In other words, heavy flavours are mainly composite states whereas light flavours are mainly elementary states. Due to the fact that SM fermions arise as admixtures of the elementary and the corresponding composite bound states, there are necessarily accompanying heavy excitations with the same SM quantum numbers. In case of the top quark, they are called (fermionic) top partners. Those top partners belong to a representation of the unbroken subgroup ${\mathcal H}$, suggesting the existence of other new particles in the same multiplet (we denote them as ``top partners'' as well). For example, if ${\mathcal H} = SO(4)$ we expect exotic top partners with electric charge 5/3 and 2/3, $X_{5/3}$ and $X_{2/3}$ respectively, in addition to a bottom-like top partner $B$~\cite{Agashe:2006at,Contino:2006qr}. Due to its exotic charge $X_{5/3}$ is supposed to be the lightest top partner of the multiplet, as it cannot mix with any SM particle.

\begin{figure}
\begin{center}
\epsfxsize=0.43\textwidth\epsfbox{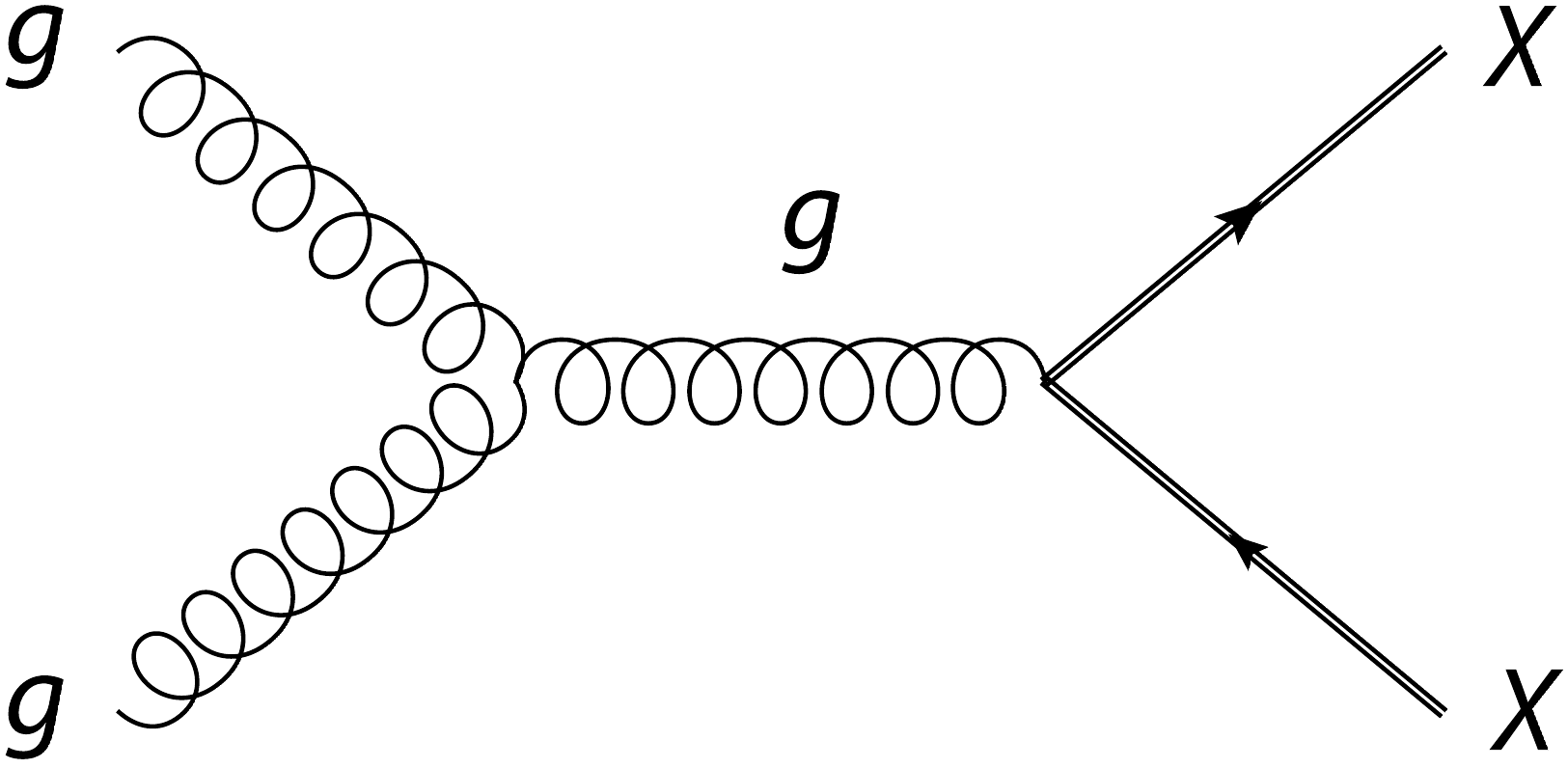}\hspace{5mm}
\epsfxsize=0.43\textwidth\epsfbox{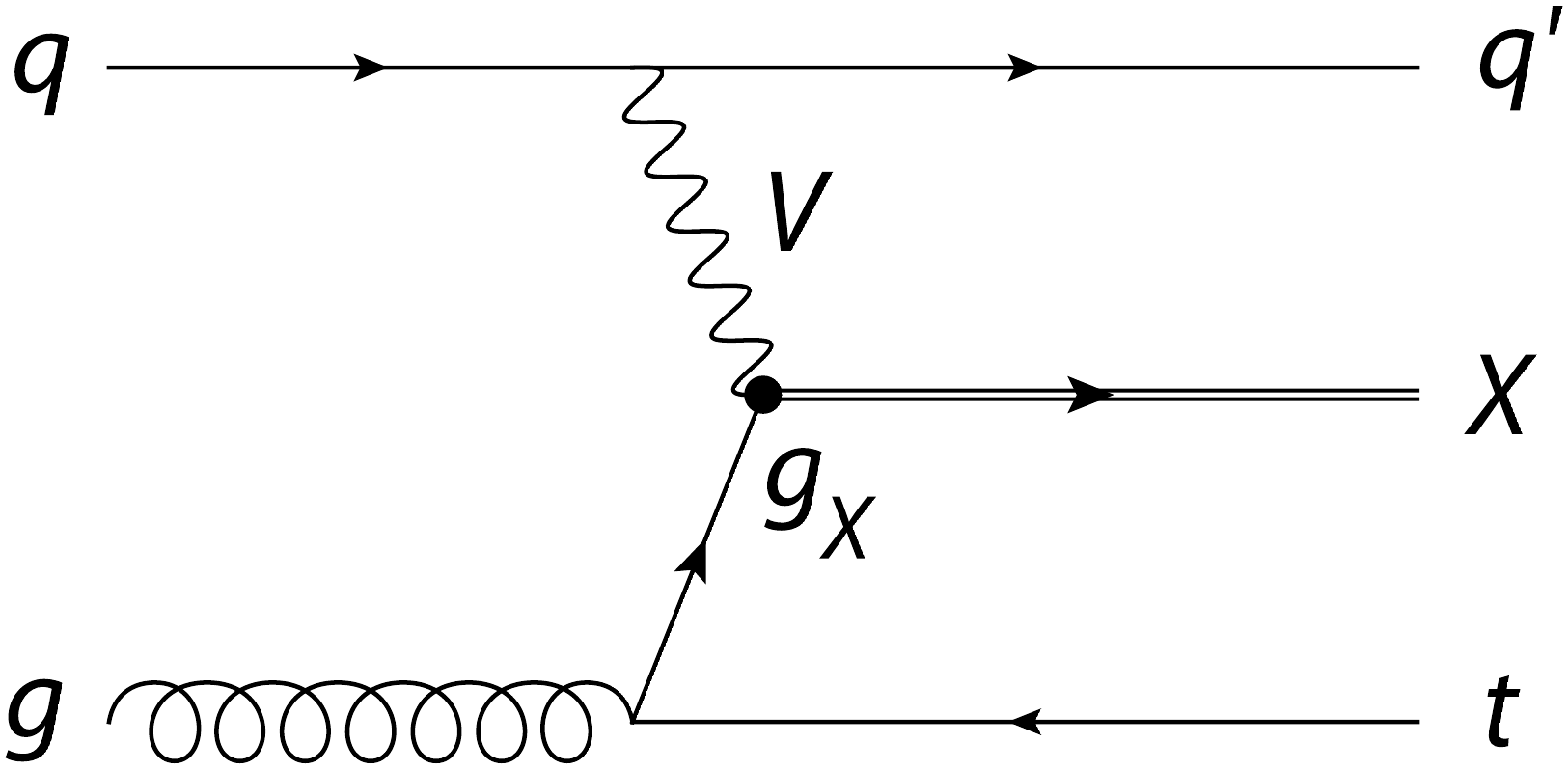}
\caption{The diagrams of the double production and single production processes. We show only the dominant diagrams. $t$ is a collective symbol to refer to either top or anti-top. Similarly $X$ denotes any top partner of interest, and it collectively refers to either top partner or anti-top partner.}
\label{fig:singleANDdouble}
\end{center}
\end{figure}

While a lot of effort has already been dedicated to setting limits on top partners in a model-independent way, we take a rather different approach. We customise the search strategy specifically for the lightest exotic top partner $X_{5/3}$. 
At the LHC the top partners can be either pair produced, through QCD interactions alone, or singly produced, involving a model dependent coupling. We show the relevant diagrams in Fig.~\ref{fig:singleANDdouble}. The relevant coupling for the single production process is~\footnote{There are also interactions involving the bottom quark, $\bar{X}{\slashed V} b$. However, we will not consider those interactions in our work.}
\beq \label{eq:singleVertex}
 g_X\, \bar{X}{\slashed V} t~,
\eeq
where $V$ denotes an electroweak gauge boson ($W^{\pm},\ Z$) and $X$ can be any top partner. The model dependence of the single production process is encoded in the couplings $g_X$ which are functions of the model parameters that determine the mass spectrum and the particles' interactions. However, we emphasise that the model-dependense we are exploiting is rather minimal, and as such required in a large class of composite Higgs models. The relative size of the two production processes entirely depends on the top partners' masses and the couplings $g_X$ probed in the single production process. Compared to the pair production process the single production cross section suffers from the exchange of a virtual electroweak gauge boson and a gluon splitting to heavy quarks. This results in a larger pair production cross section if the top partner is light. However, for heavier top partners the mainly gluon-induced pair production cross section drops quickly and the single production mode begins to take over.

While the prospects for the single production process at the LHC are discussed in~\cite{Mrazek:2009yu,Cacciapaglia:2011fx,Vignaroli:2012nf,Vignaroli:2012sf,Li:2013xba,DeSimone:2012fs}, most experimental top partner searches are based on the pair production process because of its dominant signal rates at lower masses and promising kinematic features to overcome the large Standard Model backgrounds (see~\cite{ATLAS-CONF-2013-051,ATLAS-CONF-2013-056,ATLAS-CONF-2013-060,ATLAS:2013ima,CMS:2013sin,CMS:vwa,CMS-PAS-B2G-12-021,CMS-PAS-SUS-12-027} for recent analyses on top partners using LHC8 data). Based on those searches, currently the limit on top partner masses is close to or has already passed the crossing point of the single vs double production cross section. For masses much heavier than the top quark the top partner's decay products are necessarily boosted. While this is a general feature when a heavy resonance decays into two much lighter resonances, the impact of the related kinematics on the top partner searches is dramatic.\footnote{A top partner search exploiting boosted techniques in the single production process was studied in~\cite{Li:2013xba}.}

In the mass range of interest, the increasing relevance of the single production process extends to an increased sensitivity on the model parameters\footnote{In the pair production process the model parameters can be constrained by measuring the branching fractions of top partner decays. In case the top partner decays to certain final states with unit coupling, it becomes rather insensitive to the model parameters that define the structure of the composite Higgs model.}. This fact and its implications on the minimal composite Higgs scenario were recently discussed in detail in Ref.~\cite{DeSimone:2012fs} where the same-sign dilepton and trilepton searches performed by CMS and ATLAS were recast. The authors of~\cite{DeSimone:2012fs} propose a tailored search for the single production process exploiting forward tagging jets. While this search can enhance the sensitivity to specific parts of the parameter space, we take a different path in this work. We keep our search as inclusive as possible to both the single- and pair-production processes. By combining both signal processes we enhance the total signal rate while maintaining a good statistical significance. In other words, we correlate the rates of several production mechanisms to increase the sensitivity on the model parameters ({\it process correlation}). Another promising approach we advertise to constrain the model further is to correlate the contributions of several resonances to the same search ({\it particle correlation}). The number of combined contributions to the same final state from different top partners crucially depends on their mass spectra and branching fractions which are determined by the same set of model parameters that determine the couplings $g_X$ of the single production process. If interference is negligible, the total rate of the two types of processes (single and double production) for each top partner (if multiple candidates exist) can be expressed as,
\beq \label{eq:combine}
  N_{\rm total} = \sum_X \left ( N^X_{\rm pair} + g_X^2\, N^X_{\rm single} (g_X=1) \right ).
\eeq
$N$ denotes the number of events for a given integrated luminosity and the index $X$ runs over the top partners. To facilitate a scan over the model dependent parameter $g_X$ we explicitly factor it from the single production rate. We will argue in this work that our proposed search strategy can significantly improve the limit setting on top partner masses while simultaneously probing a large part of the model's parameter space. 

All signal final states with sufficient overlap are of potential interest for such a cross correlation. More precisely, after the decay of the two resonances $X_{5/3}$ or $B$ to $tW$, in both the single and double production processes, a $t\bar{t}W$ system will emerge, see Fig.~\ref{fig:singleANDdouble}. Naturally, a signature that can be used to disentangle the signal from the SM backgrounds is the clean final state with same-sign dileptons and jets. However, particularly at the LHC with 8 TeV center-of-mass energy an economical use of the signal rate is crucial. We will demonstrate that the final state with one lepton and jets has the potential to exceed the exclusion limits set by the same-sign dilepton search.

The absence of any excess in existing new physics searches leads us to consider heavier top partners. Electroweak scale resonances, $W$/$Z$/$h$/top, decayed from those heavy top partners are necessarily boosted, and as a result all the final state particles of those boosted tops and electroweak bosons are collimated in the laboratory frame. 
As we are in a transition region between the boosted and unboosted regime, many standard search strategies cease to work well. Relatively simple observables like the multiplicity of jets proved to be a powerful discriminator between signal and SM backgrounds, yet when the decay products are collimated in a narrow opening angle overlapping radiation will spoil the aimed-for jet-parton matching and jet counting becomes much less effective. A simple example is the reconstruction of a $W$ boson. A $W$ boson with small transverse momentum decays to two widely separated jets, counted as two, whereas for a highly boosted $W$ boson, whose transverse momentum is much bigger than its mass ($p_{T,W} \gg m_W$), the two jets merge into a single jet which would be counted as one. This obscures the two-prong nature of the $W$ boson, as opposed to the one-prong QCD-jet. Here jet substructure techniques can be helpful to recover sensitivity in discriminating the $W$ jet from QCD jets. Those techniques organise the energy distribution of jet constituents such that they correctly identify two hard objects in a single jet, while efficiently rejecting QCD-like jets. By exploiting the substructure of a jet the traditional observable {\it jet multiplicity}, $N_j$, can be consistently extended to $N_{\rm con}$, the number of jet constituents\footnote{This variable is already being explored in a top partner search by CMS~\cite{CMS:vwa}.}, defined as
\beq \label{eq:Ncon}
  N_{\rm con} = \sum_{n=1} n\cdot N_{\rm n-prong}.
\eeq
$N_{\rm n-prong}$ is the number of boosted $n$-prong jets, tagged by $n$-prong taggers, and $N_{\rm 1-prong}$ is just the number of traditional jets. In this classification, top taggers~\cite{Brooijmans:2008zza,Thaler:2008ju,Kaplan:2008ie,Almeida:2008yp,Almeida:2010pa,Plehn:2010st,Ellis:2009su,Thaler:2010tr,Soper:2012pb,Schaetzel:2013vka} (see also~\cite{Abdesselam:2010pt,Altheimer:2012mn}) are 3-prong taggers and $W$/$Z$/$h$-taggers~\cite{Seymour:1993mx, Butterworth:2002tt, Butterworth:2008iy, Almeida:2008yp, Almeida:2010pa, Ellis:2009su, Krohn:2009th,Soper:2010xk,Cui:2010km,Thaler:2010tr,Hackstein:2010wk,Soper:2011cr,Almeida:2011aa,Dasgupta:2013ihk} are 2-prong taggers. A crucial advantage of this approach is that it continuously {\it interpolates between the boosted and unboosted regime}. Therefore sensitivity for the signal is restored over a maximum range in phase space.

2- and 3-prong taggers do not only count the number of subjets inside a jet but also impose kinematic requirements, e.g. the reconstruction of the correct top or W mass. Thus, we find that apart from $N_{\rm con}$ the number of reconstructed top quarks $N_{\rm top}$ and W bosons $N_{\rm W}$ using jet substructure techniques provide a strong handle in disentangling the signal from the SM backgrounds. In general top taggers do not make use of b-tagging. However, b-tagging can be applied in addition to a top tag \cite{Harigaya:2012ir}, though we will not pursue this possibility.

We will investigate a cut-and-count analysis based on $N_{\rm con}$, $H_T$, $N_{\rm W}$ and $N_{\rm top}$ in the context of top partner searches. In Section~\ref{sec:model} we begin with a brief description of our simplified model approach. In Section~\ref{sec:constraints} we recast the constraint of a recent CMS search on $X_{5/3}$ in the same-signe dilepton channel and extend it into two-dimensional exclusion regions by summing over the contributions from the accessible processes and top partners. In Section~\ref{sec:strategies} we quantify the exclusion reach for the top partners with LHC8 data. Finally we conclude with a discussion of the implications of the results from Sections~\ref{sec:constraints} and \ref{sec:strategies}.

\section{A Simplified Model}
\label{sec:model}

%%%%%%%%%%%%%%%%%%%%%%
%  A Simplifed Model
%%%%%%%%%%%%%%%%%%%%%%
For the discussion of LHC phenomenology of top partners we consider a simplified model of the minimal composite Higgs scenario based on $SO(5)/SO(4)$~\footnote{While the simplified top partner models of generic composite Higgs scenarios have been discussed in~\cite{Contino:2006nn,Contino:2008hi,Mrazek:2009yu}, the interpretation of our result in those models is straightforward as the couplings and mass spectrum take simple forms.}. We assume that $t_R$ is a completely composite chiral state and only $q_L = (b,\ t)_L$ is realised as partially composite. To avoid large corrections of the $T$-parameter and a large $Zb_L\bar b_L$ interaction we assume the custodial symmetry is preserved~\cite{Agashe:2006at}. For simplicity we consider the case where top partners belong to the fourplet, $\Psi \equiv {\bf 4_{2/3}}$, embedded in ${\bf 5_{2/3}}$ of $SO(5)$. The four components of the fourplet are $T,B,X_{2/3},X_{5/3}$. Here $2/3$ refers to the charge of an extra $U(1)_X$, introduced for the correct assignment of the SM hypercharge $Y=T^3_R+X$. LHC phenomenology for those top partners has been discussed in~\cite{DeSimone:2012fs}. The leading order Lagrangian in Callan-Coleman-Wess-Zumino (CCWZ)~\cite{Coleman:1969sm,Callan:1969sn} language is given by
\beq
\begin{split} \label{eq:M54}
  {\cal L} =& \ {\cal L}_{\rm kin} - \bar{\Psi} {\slashed e} \Psi - M_{\Psi}\bar{\Psi}\Psi \\
&+ i\, c_1 (\bar{\Psi}_R)_i \gamma^\mu d^i_\mu t_R + y\,f\, (\bar{Q}^{\bf 5}_L)^I U_{I\, i} \Psi^i_R + y\, c_2\, f (\bar{Q}^{\bf 5}_L)^I U_{I\, 5}t_R + {\rm h.c.}~,
\end{split}
\eeq
where ${\cal L}_{\rm kin}$ includes the covariant kinetic terms of $q_L,\, t_R\,$ and $\Psi$ (see~\cite{DeSimone:2012fs} for details about the conventions used in Eq.~\ref{eq:M54}). The pNGB Higgs is parameterised by the matrix $U$ in Eq.~\ref{eq:M54}, defined as $U \equiv {\rm exp}[i\sqrt{2}/f\ \Pi_i T^i]$. $T^{i=1\cdots 4}$ are broken generators parameterising the coset of $SO(5)/SO(4)$ and $f$ is the associated symmetry breaking scale. The simplified model in Eq.~\ref{eq:M54} leads to a non-tunable structure of the Higgs potential at leading order (see~\cite{Panico:2012uw} for other possibilities). While top partners can be embedded in a bigger representation of $SO(5)$~\cite{DeSimone:2012fs} or be part of a less minimal model, the physics captured by the simplified model in Eq.~\ref{eq:M54} is likely to be a subset of them. The model is defined by five parameters in addition to the ones of the SM. One of them is fixed by the top mass constraint, leaving eventually only four free parameters

First of all, $X_{5/3}$ can not mix with any other state due to its exotic charge. Hence its leading-order physical mass is expressed by $M_{\Psi}$ in Eq.~\ref{eq:M54}. The bottom-type quark sector takes the form of a $2 \times 2$ mass matrix in the basis of $(b,\ B)$, and the mass of the heavy eigenstate is derived from Eq.~\ref{eq:M54} as
\beq \label{eq:Bmass}
  m_B = \sqrt{M^2_{\Psi} + y^2f^2}~.
\eeq
Eq.~\ref{eq:Bmass} holds even after electroweak symmetry breaking. The mass spectrum of the up-type quark sector can be obtained by diagonalising the $3\times 3$ mass matrix in the basis of $(t,\ T,\ X_{2/3})$. It was pointed out in Ref.~\cite{DeSimone:2012fs} that the mass matrix can be made to be block-diagonal, by an appropriate field redefinition of $T$ and $X_{2/3}$, such that the mass of $X_{2/3}$ can be expressed by $M_{\Psi}$. In other words, in this model the masses of $X_{2/3}$ and $X_{5/3}$ are degenerate. The masses with non-trivial dependence on the model parameters are  those of $t$ and $T$. They are approximately given by~\footnote{Note, in our numerical evaluations in Secs.~\ref{sec:constraints} and \ref{sec:strategies}, we use exact expressions.}
\beq
\begin{split} \label{eq:topmass}
  m_t \sim &\quad \frac{c_2\, y f}{\sqrt{2}}\frac{g_{\Psi}}{\sqrt{g^2_{\Psi} + y^2}} \sqrt{\xi} \left [ 1 + {\mathcal O} \left (\frac{y^2}{g^2_\Psi} \xi \right ) \right ]~,\\[1pt]
  m_T \sim &\quad \sqrt{M^2_\Psi + y^2 f^2} \left [ 1 - \frac{y^2 (g^2_\Psi + (1-c^2_2)y^2)}{4 (g^2_\Psi + y^2)^2}\, \xi + \cdots \right ]~,
\end{split}
\eeq
where $g_\Psi \equiv M_{\Psi}/f$ and $\xi \equiv (v/f)^2$. 

The top partners $X_{5/3}$ and $B$ are in this model particularly interesting. The $X_{5/3}$ is the lightest top partner and it decays to $tW$ with unit coupling\footnote{In a more general setup, this property can be relaxed such that $X_{5/3}$ also decays to $Wq$ ($q=u,c$)~\cite{Cacciapaglia:2012dd}.} due to its exotic charge. If this type of composite Higgs model is indeed responsible for electroweak symmetry breaking and is kinematically accessible at the LHC, $X_{5/3}$ is the most promising candidate to be discovered first. The mass limit on $X_{5/3}$ automatically extends to other top partners via the mass relations of Eqs.~\ref{eq:Bmass} and \ref{eq:topmass}, constraining all top partners indirectly. For instance, while the search for $X_{2/3}$ is more difficult due to its predominant decay to $tZ$ or $th$, and their subsequent mostly hadronic decays, one can get a stringent bound from the limit on $X_{5/3}$. According to Eqs.~\ref{eq:Bmass} and~\ref{eq:topmass} the mass hierarchy between $T$, $B$ and $X_{5/3}$ is controlled by $yf$. By measuring the masses of $X_{5/3}$ and $B$ the overall mass scale of the top partners $M_\Psi$ and the symmetry breaking scale $f$ can be constrained simultaneously. Remarkably, both states decay predominantly into the same class of particles\footnote{In the composite Higgs model, defined by Eq.~\ref{eq:M54}, the branching fraction of $B$ to $tW$ is dominant as $th$ and $tZ$ modes are forbidden~\cite{DeSimone:2012fs}.}: a top quark and a W boson. Therefore, the same search strategy can be applied to both particles simultaneously irrespective of whether they are singly or pair produced. When including contributions from both particles in one search, the reduced signal rate for the heavier $B$ can be compensated by the contribution from $X_{5/3}$, assuming $m_{X_{5/3}}$ not being far from $m_B$, such that a sizeable total signal rate is maintained.
Therefore, in the following we will focus on the two most accessible top partners\footnote{While there can be an extra contribution from $X_{2/3}$ to the one-lepton channel when decaying to $tZ$, the branching fraction of $tZtZ$ to one lepton is roughly two times smaller than for the $tWtW$ system. Further, BR($X_{2/3}\rightarrow tZ)\simeq 0.5$ in this model.} $X_{5/3}$ and $B$.

The single production process is directly sensitive to the details of the strong dynamics in our simplified model. The coupling in Eq.~\ref{eq:combine} is derived from the model's free parameters,
\beq \label{eq:singleVertexM54}
 g_X = g_X(f,\, y,\, c_1,\, c_2,\, M_\Psi)~,
\eeq
where one of the parameters, for instance $c_2$, can be removed using the top mass constraint. $y$ controls the mixing between elementary and the composite states, which leads to the partial compositeness of the left-handed top and bottom quarks. $y$ explicitly breaks the $SO(5)$ symmetry, generating a leading contribution to the Higgs potential. $c_2$ is expected to be ${\mathcal O}(1)$. Together with other parameters it sets the top mass. $c_1$ is also expected to be ${\mathcal O}(1)$ and it constitutes the dominant interaction for the single production process with an associated top. 
While we use the exact expressions for the coupling constants in Eq.~\ref{eq:singleVertex} in the unitary gauge for numerical evaluation, the dominant contribution to the coupling in Eq.~\ref{eq:singleVertexM54} can be estimated by the Goldstone boson equivalence theorem. For instance, the vertices of $\phi^+\bar{X}_{5/3L}t_R$ and $\phi^- \bar{B}_L t_R$ using mass eigenstates are~\cite{DeSimone:2012fs}
\beq
\begin{split} \label{eq:verticesGBET}
    &g_{X_{5/3}} \sim \sqrt{2}\, c_1 \frac{M_{\Psi}}{f}~,\\
    &g_B \hspace{4mm} \sim \sqrt{2}\, c_1 \sqrt{y^2 + (M_{\Psi}/f)^2} - c_2 \frac{y^2}{\sqrt{y^2 + (M_\Psi/f)^2}}~. 
\end{split}
\eeq
The $\phi^\pm$ are the Goldstone bosons eaten by $W^\pm$. The couplings in Eq.~\ref{eq:verticesGBET} are basically Yukawa couplings of the composite states, implying that the single production processes can be sizeable.

\section{Reinterpretation of existing searches}
\label{sec:constraints}

%%%%%%%%%%%%%%%%%%%%%%
%  Constraints
%%%%%%%%%%%%%%%%%%%%%%

\begin{figure}
\begin{center}
\epsfxsize=0.48\textwidth\epsfbox{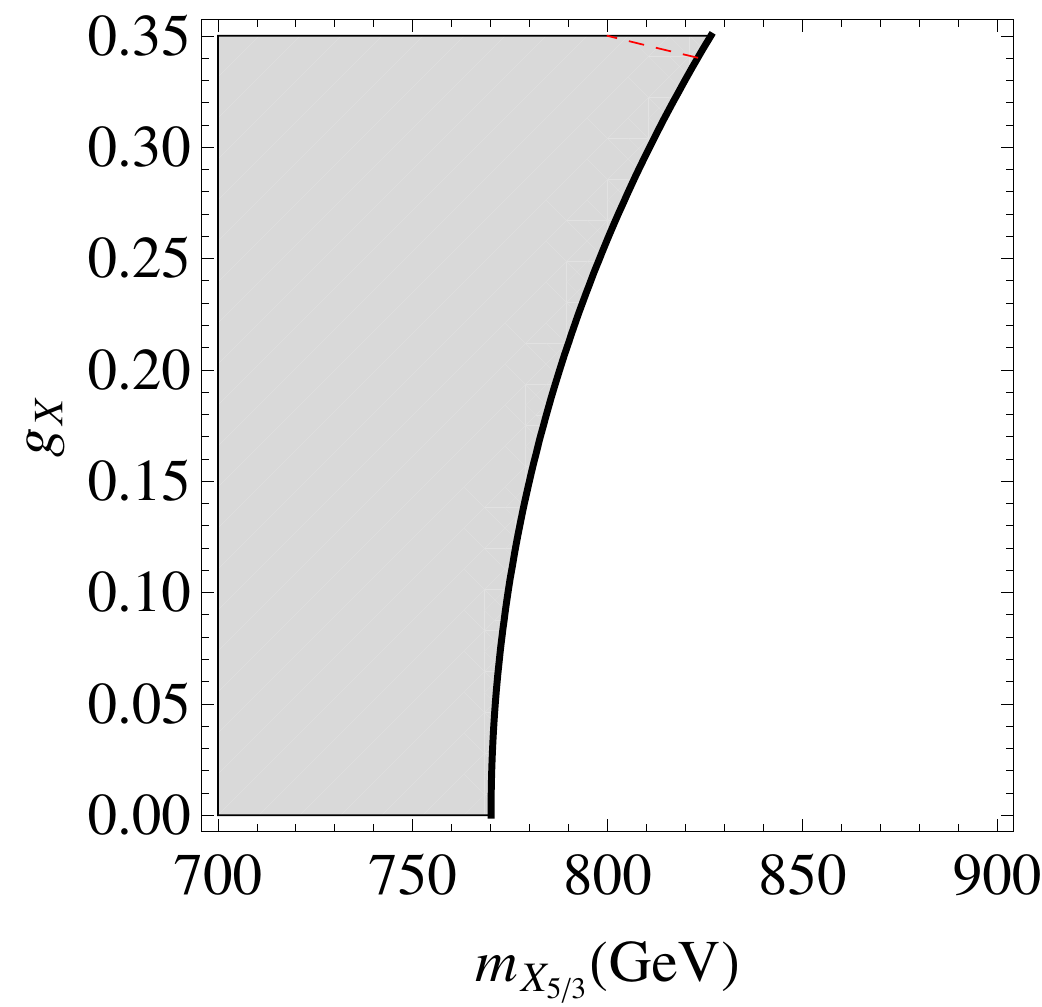}
\epsfxsize=0.48\textwidth\epsfbox{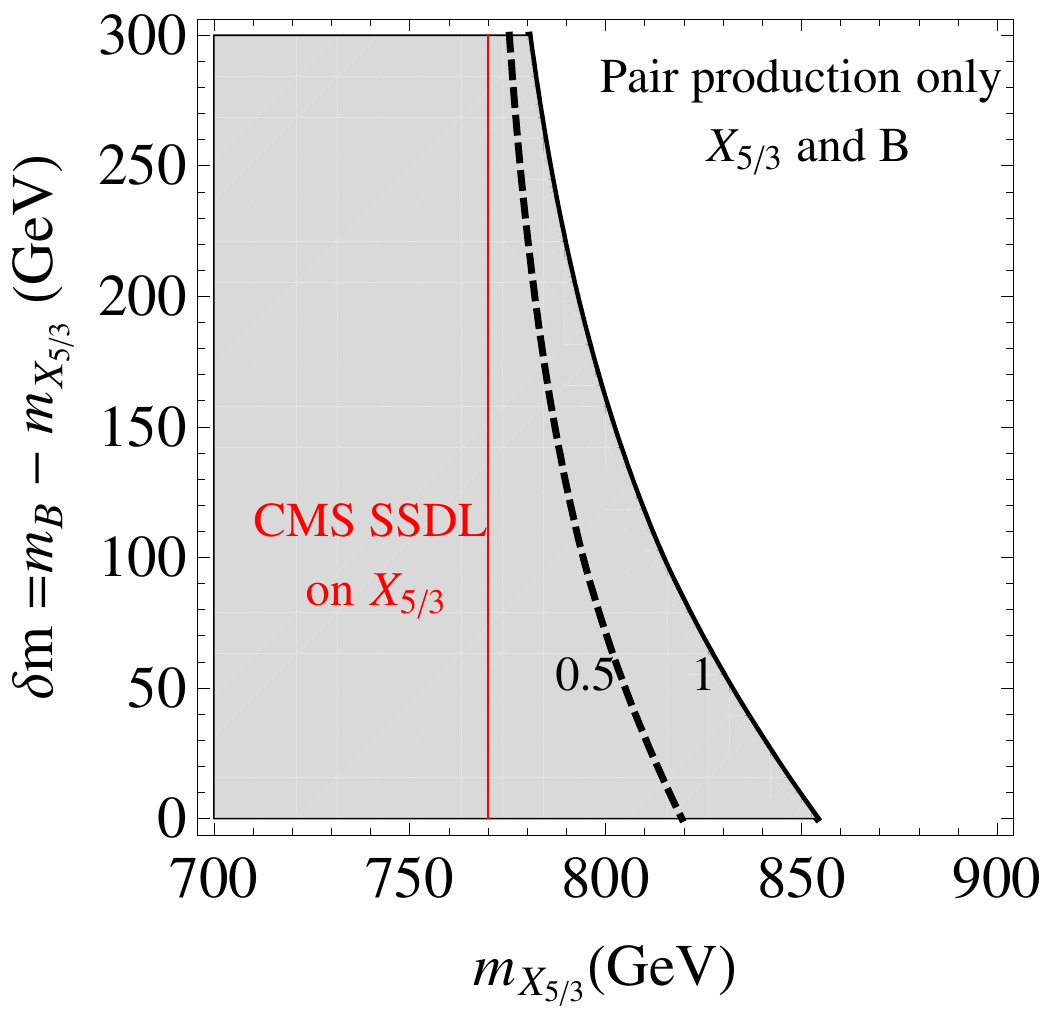}
\caption{Recasted CMS SSDL search. Left: the single production process is added such that the limit on the mass extends into a two-dimensional exclusion $(m_X,\, g_X)$-plane. Right: the contribution from the bottom-like top partner $B$ is added to those of $X_{5/3}$. Plot is restricted to only pair production processes of $X_{5/3}$ and $B$ for simplicity. BR($X_{5/3}\rightarrow tW$)=1 is assumed, and BR($B\rightarrow tW$)=0.5 (dashed black) and 1 (solid black) are plotted. Red line in the right panel indicates the bound on $X_{5/3}$ set by CMS SSDL.}
\label{fig:gVSmassCMSSSDL}
\end{center}
\end{figure}

Due to the importance of top partners for the naturalness problem, their search is part of the core program of ATLAS and CMS. In this section, with Eqs.~\ref{eq:Bmass}-\ref{eq:singleVertexM54} in mind, we will discuss the impact of existing searches, performed by ATLAS and CMS, on $m_{X_{5/3}}$ and $m_B$. We will demonstrate that recasting searches in terms of a combined measurement of $X_{5/3}$ and $B$ in the single and double production channels, as outlined in Eq.~\ref{eq:combine}, directly results in an improved limit on the simplified model's parameters.

The four processes we exploit to perform this task are $pp \to X_{5/3} \bar{X}_{5/3} \to W^+ t W^- \bar{t}$, $pp \to B \bar{B} \to W^- t W^+ \bar{t}$, $pp \to X_{5/3} \bar{t} + {\rm h.c.} \to W^+ t \bar{t}$ + h.c. and $pp \to \bar{B} t + {\rm h.c.} \to W^+ \bar{t} t$ + h.c. All of the processes can give rise to same-sign (SS) dilepton signatures.

CMS recently published a search~\cite{CMS:2013sin} which explores many channels that include a varying number of leptons (i.e. one lepton, SS dileptons, two types of opposite-sign (OS) dileptons and trileptons), to derive the limits for various final states, e.g. $bW$, $tZ$, $tH$, in pair produced vector-like top partner signals. While their SS dilepton search will pick up our signals, a more tailored SS dileption search targeted on $X_{5/3}$, using full LHC8 data, has been shown in~\cite{CMS:vwa}. To date this search yields the strongest bound on the top partner $X_{5/3}$ using 19.6 fb$^{-1}$ of LHC8 data. The limit is set using the pair-produced $X_{5/3}$, each of them decaying to $tW$ with unit branching fraction. This search not only demands same sign dileptons, but also exploits jet substructure techniques to efficiently capture boosted top quarks or $W$ bosons. The $W$ can either originate from the top or directly from $X_{5/3}$. The search imposes the cuts $H_T > 900$ GeV and $N_{\rm con} \geq 5$, resulting in a limit of $m_{X_{5/3}} > 770$ GeV at 95\% CL.

\begin{table}[tbp]
\centering
\begin{tabular}{ccccc}  
\hline
 \quad $X_{5/3}$ Mass (GeV)  & \quad 2SS leptons  & \quad $m(ll)$ Veto & \quad $N_{\rm con} \geq 5$ & \quad $H_T \geq 900$ \quad \\
\hline \hline
\multicolumn{5}{c}{Pair production of $X_{5/3}$ ($\rightarrow tW$)}\\
 700  & 55.7 & 49.5 & 29.4 & 20.6  \\
 800  & 20.1 & 18.3 & 11.6 & 9.48  \\
 900  & 7.88 & 7.22 & 4.59 & 4.08  \\
1000  & 3.33 & 3.1 & 2.01 & 1.89  \\
\hline
\multicolumn{5}{c}{Single production of $X_{5/3}$ ($\rightarrow tW$) with an associated $\bar t$}\\
 700  & 579.5 & 545.8 & 146.9 & 45.4 \\
 800  & 388.6 & 365.8 & 101.1 & 38.8 \\
 900  & 250.9 & 234.4 & 65.1  & 30.4 \\
1000  & 173.8 & 163.4 & 48.9  & 26.3 \\
\hline \hline
\end{tabular}
\caption{Summary table of the expected signal events for the pair production (as the validation of our analysis) and the single production processes. Branching fraction of the pair production (single production) to same-sign (SS) leptons is 0.21 (0.11). The expected signal events of the single production process assumes $g_X = 1$.}
\label{tab:CMSSSDL}
\end{table}

Starting from the findings of \cite{CMS:vwa} we can improve the limit and extract more information on the model parameters. While the cuts in this search are rather exclusive to the doubly produced $X_{5/3}$, a significant fraction of events where $X_{5/3}$ is produced in association with a top quark still passes those cuts, hereby contributing to the total signal rate. We show the number of signal events passing the different event selection cuts in Table~\ref{tab:CMSSSDL}. By combining the single and double top partner production according to Eq.~\ref{eq:combine}, the separate limits on either the model dependent coupling $g_X$ or the top partner mass $m_{X_{5/3}}$ can now be unfolded on a two-dimensional exclusion plane. 

We run the same analysis on the signals of the single and pair production processes, generated by {\tt MadGraph5} {\tt v1.4.7}, interfaced with our own simplified UFO~\cite{Degrande:2011ua} model, processed through {\tt PYTHIA8} and clustered using {\tt FastJet} {\tt v3.0.3}~\cite{Cacciari:2005hq}. We normalise signal cross sections to their NNLO values, derived by {\tt HATHOR}~\cite{Aliev:2010zk}. We validate our procedure by comparing the pair production signals with the CMS results, see Table~\ref{tab:CMSSSDL}. In Fig.~\ref{fig:gVSmassCMSSSDL} we show that the limit on $m_{X_{5/3}}$ can be improved to $\sim$ 830 GeV for $g_X=0.35$ this way. As discussed in Sec.~\ref{sec:model} the coupling $g_X$ is a function of the model parameters which determine the mass spectrum of the top partners. Therefore, an extension of the exclusion region for $m_{X_{5/3}}$ in comparison with the limit from pair production alone constitutes a strong discriminator between the models.

So far we only studied the impact of the combination of the single and double $X_{5/3}$ production processes on the exclusion plane. As mentioned in Sec.~\ref{sec:model}, by including the bottom-like top partner $B$ we can access orthogonal information on the model parameters. As long as the mass hierarchy between $B$ and $X_{5/3}$ is small, they can both contribute to the same final state with a sizeable rate. We only consider the double production processes of $X_{5/3}$ and $B$. The samples of $B$ are simulated and processed in the same way as the ones for $X_{5/3}$. While the inclusion of $B$ introduces two more free parameters, namely its mass $m_B$ and the branching ratio to $tW$, we demonstrate this effect on the limit of $m_{X_{5/3}}$ for a varying mass gap $\delta m \equiv  m_B - m_{X_{5/3}}$ with two different choices of the branching ratio. We show the results in the right panel of Fig.~\ref{fig:gVSmassCMSSSDL}. 

One can further include the single production processes of $X_{5/3}$ and $B$, hereby introducing sensitivity on two more couplings. We will discuss this option  in  Section~\ref{sec:conclusions} in more detail.

In addition to the SS leptons, the one lepton channel is also very sensitive to our signal. Multivariate techniques are used in \cite{CMS:2013sin} , i.e. a boosted decision tree, to infer the limit from the one lepton final state. The reported expected (observed) limit for BR($tZ$)=1 is 689 (644) GeV. If one applies this search to the pair produced $X_{5/3}$ with subsequent decay to $tW$, the limit can be even stronger due to the roughly two times larger signal rate. While we do not attempt to recast the search channels in~\cite{CMS:2013sin} to give a reinterpretation in terms of the ($m_X$, $g_X$) exclusion plane, performing such an analysis would be straightforward. Instead we will discuss the one lepton search in the context of a different search strategy in Section~\ref{sec:strategies}.

\section{Boosting searches using jet substructure}
\label{sec:strategies}

%%%%%%%%%%%%%%%%%%%%%%
%  Reconstruction techniques
%%%%%%%%%%%%%%%%%%%%%%

While for our signal processes final states with same-sign dileptons and trileptons are certainly the cleanest to exploit, in this Section we will focus on the one lepton channel. By changing from the SS dileptons, discussed in Sec.~\ref{sec:constraints}, to the one lepton channel we increase the rate in the pair production process by a factor six and in the single production process by a factor nine. Consequently the relative sensitivity of the single production process with respect to the pair production process is increased as well. A lepton in the one lepton channel can be produced from either of the $W$s. The other two or three $W$s will decay hadronically with accompanying b-jets if they originate from top quarks. The hadronic tops and $W$s from the heavy top partner's decay, as opposed to those in SM backgrounds, are necessarily boosted, while the ones produced in association with the top partner (in the single production process) are not. There is only one source of missing transverse momentum in the final state. Therefore by requiring the lepton and the neutrino to reconstruct the W mass $M_W = M_{\nu l}$ one can reconstruct the leptonic top. This way we are able to reconstruct the entire $ttW$ subsystem, from which we can reconstruct the top partner mass.

The single and double production processes, both for $X_{5/3}$ and $B$, share a common feature: they contain a $ttW$ subsystem which can lead to a final state with one lepton and at least six partons. This ensures that the signal will likely populate the high-valued $N_{\rm con}$ region. The $ttW$ subsystem of the signal is produced from the decay of heavy top partners, as opposed to those from non-resonant QCD processes, and thus the $p_T$-summed $H_T$ is roughly proportional to the heavy top partner masses. Thus, $N_{\rm con}$ and $H_T$ are very effective observables to separate the signals from backgrounds.\\

The signal samples are simulated as in Sec.~\ref{sec:constraints}. The major backgrounds in our one lepton analysis include $t\bar t$+jets matched up to two jets and $W$+jets matched up to four jets using MLM matching~\cite{Alwall:2007fs} with $R = 0.4$ and $p_T = 30$~GeV. Conservatively, we apply a $K$-factor of 2 to both backgrounds. The events are not further processed to take into account detector effects. We check other irreducible backgrounds such as $t\bar{t}W$+jets, and we find that those are subleading. 

%--------------------------------------
\subsection{Cut and count analysis for $l$+jets final state}
\label{subsec:cutandcountljets}
%--------------------------------------
\begin{figure}
\begin{center}
\epsfxsize=0.45\textwidth\epsfbox{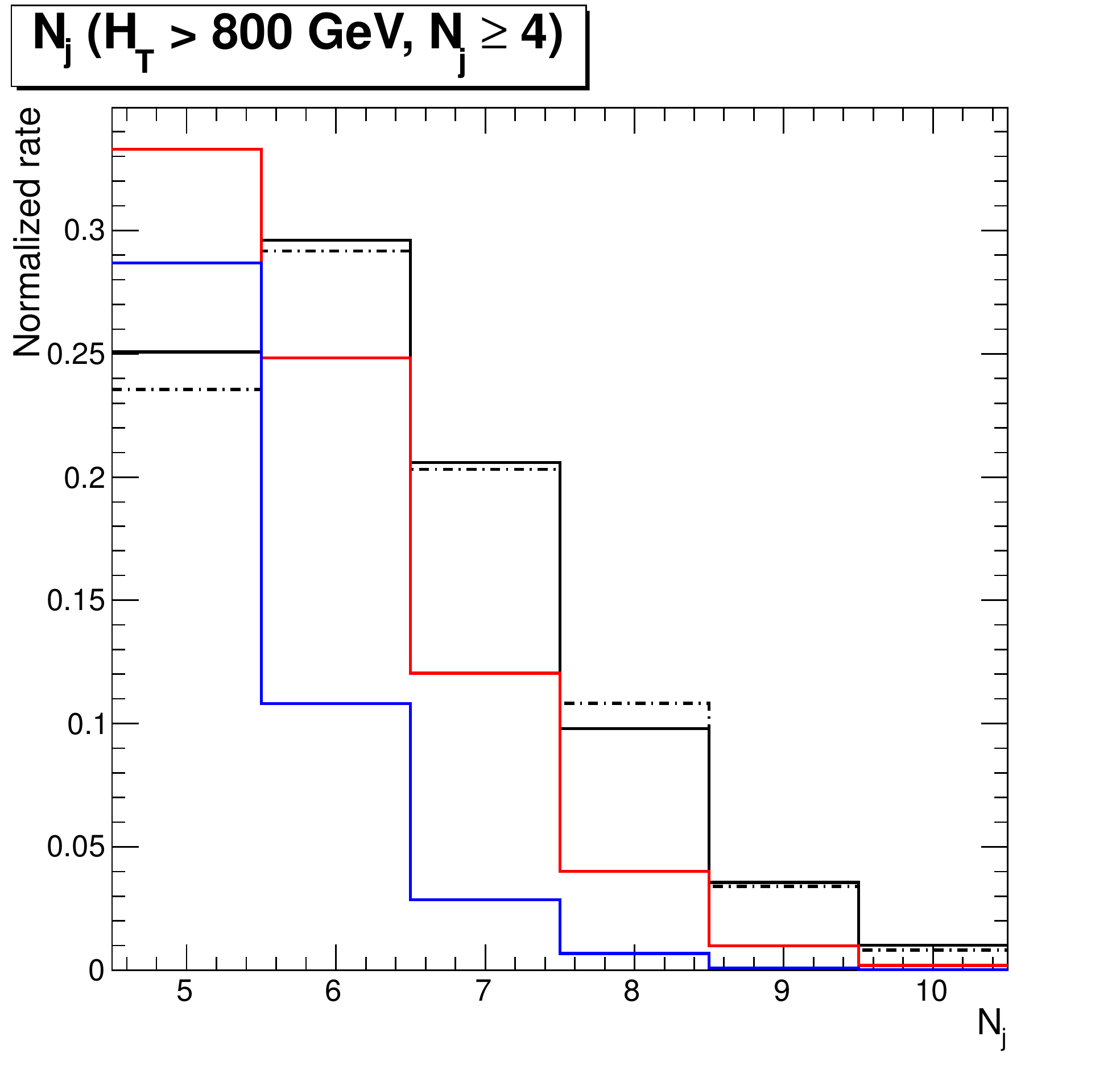}
\epsfxsize=0.45\textwidth\epsfbox{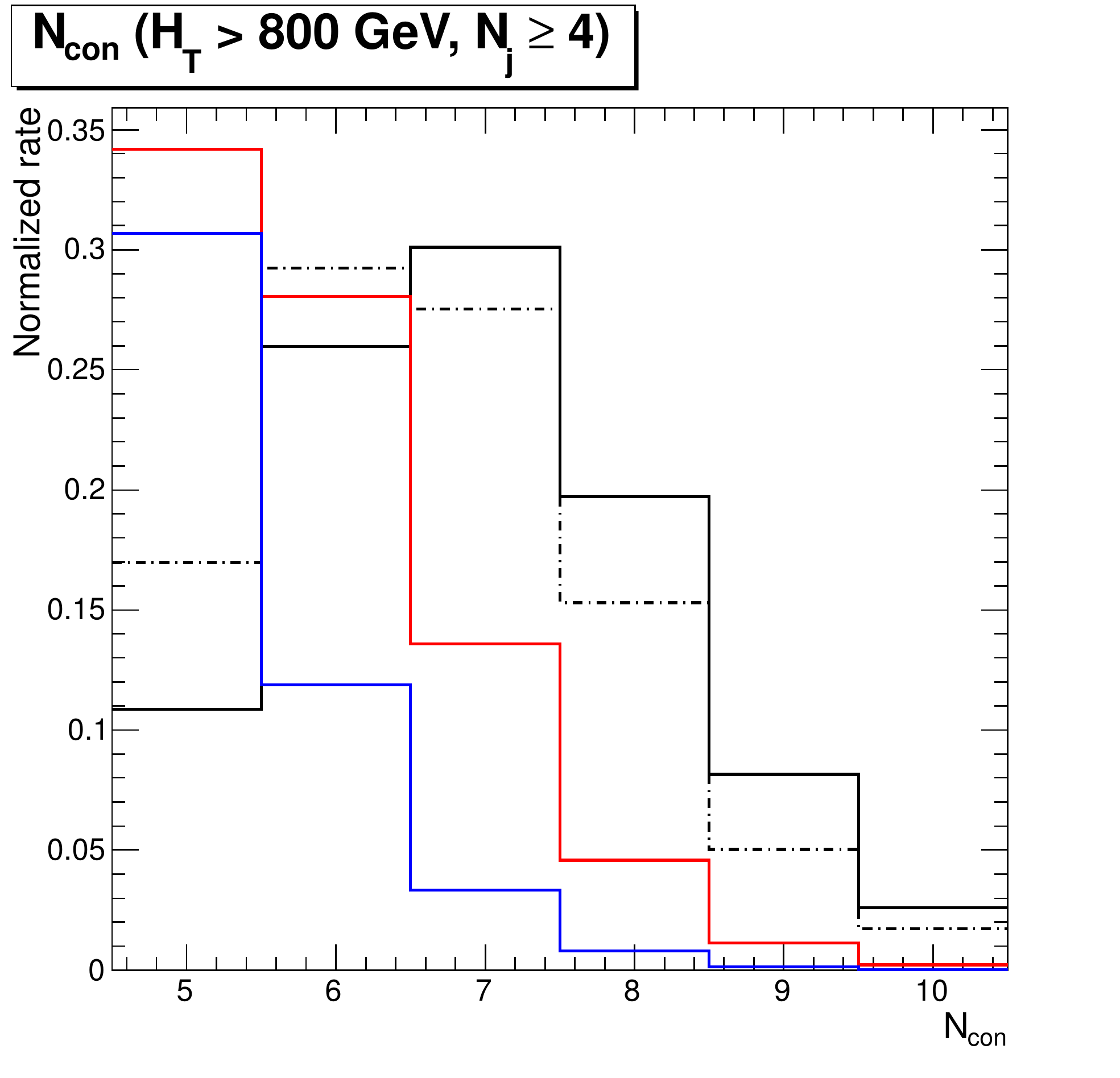}
\caption{Normalized distributions of the number of traditional jets (left) and of the constituents (right) for signals and backgrounds in $l$ + jets channel. Signals are the pair (black solid) and single (black dashed) production processes of 800 GeV $X_{5/3}$, and the backgrounds are $t\bar t$ + jets (solid Red) and $W$ + jets (solid blue). Events in the plots were restricted to those satisfying $H_T > 800$ GeV and $N_j \geq 4$. The area of each curve over the full range of $N_j, N_\mathrm{con}$ is normalized to 1. We only display $5 \leq N_j,N_\mathrm{con} \leq 10$.}
\label{fig:NjandNconOneLepton}
\end{center}
\end{figure}

\begin{figure}
\begin{center}
\epsfxsize=0.45\textwidth\epsfbox{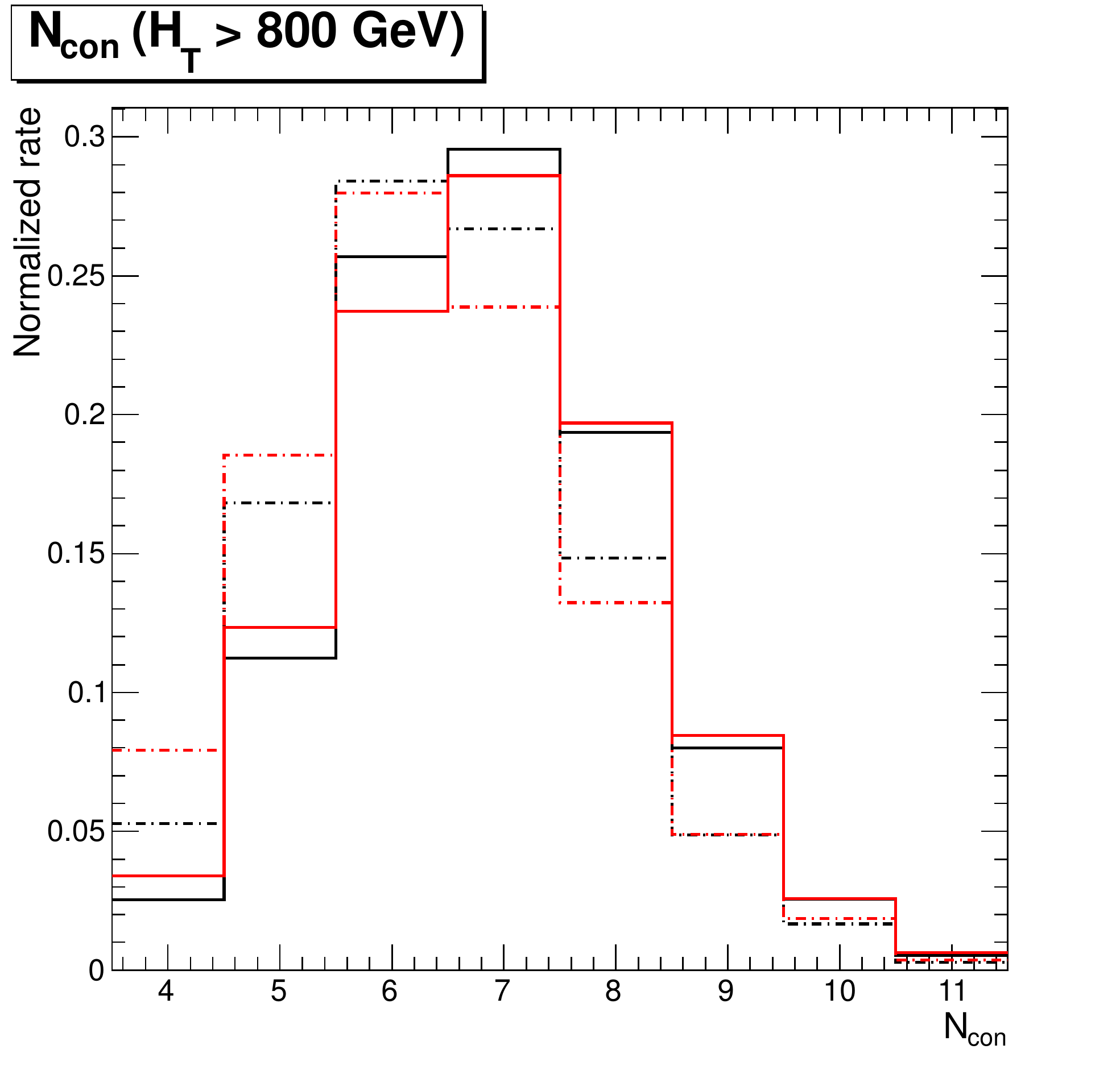}
\epsfxsize=0.45\textwidth\epsfbox{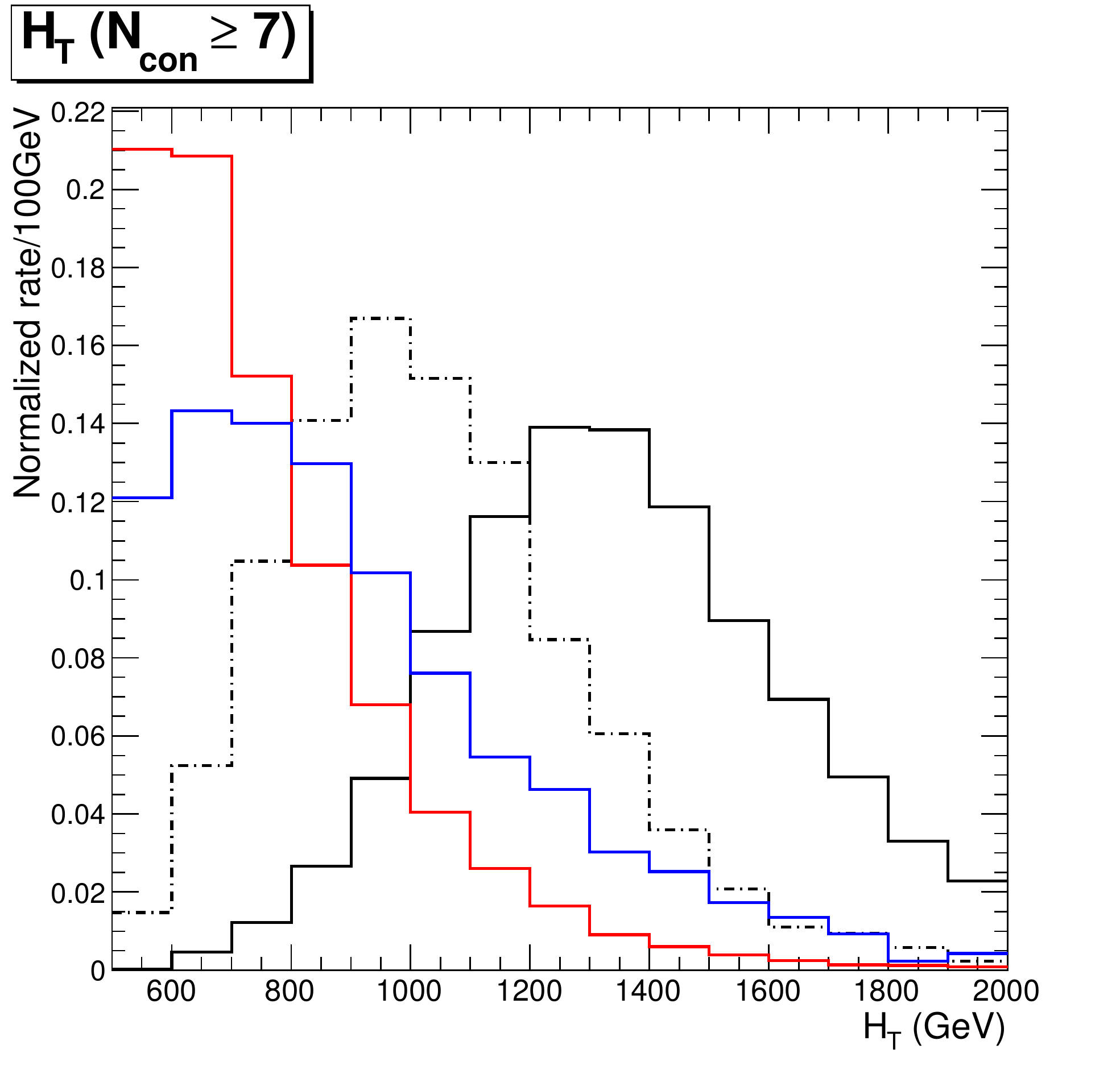}
\caption{Left: Number of reconstructed constituents for the $X_{5/3}$ top partner with 800 (black) and 1000 (red) GeV from pair (solid) and single (dashed) production processes. Right: $H_T$ distributions of the $X_{5/3}$ with 800 GeV from pair (solid black) and single (dashed black) production processes. The background distributions are $t\bar t$ + jets (solid red) and $W$ + jets (solid blue).}
\label{fig:HTandNjOneLepton}
\end{center}
\end{figure}

\begin{figure}
\begin{center}
\epsfxsize=0.31\textwidth\epsfbox{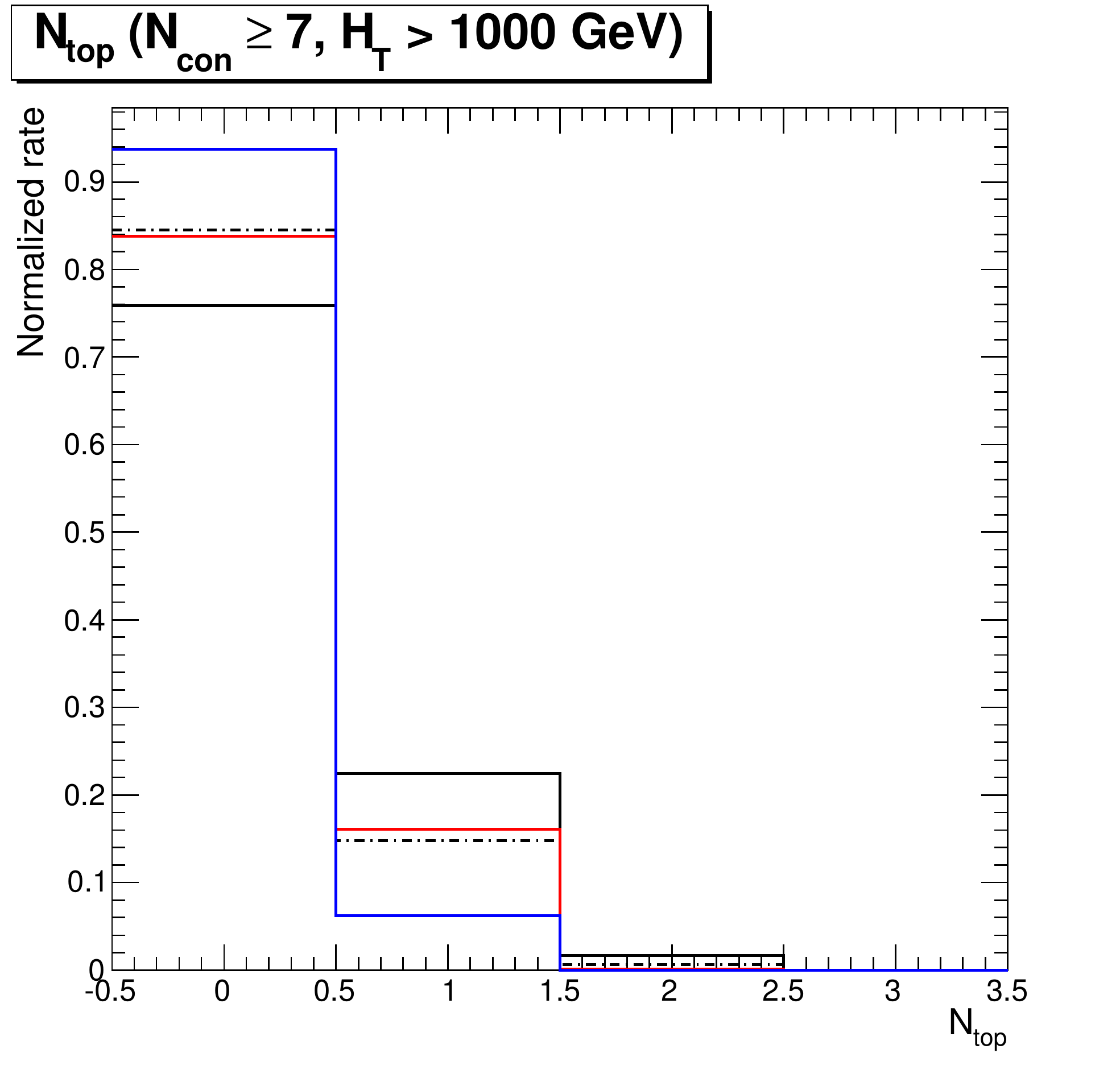}
\epsfxsize=0.31\textwidth\epsfbox{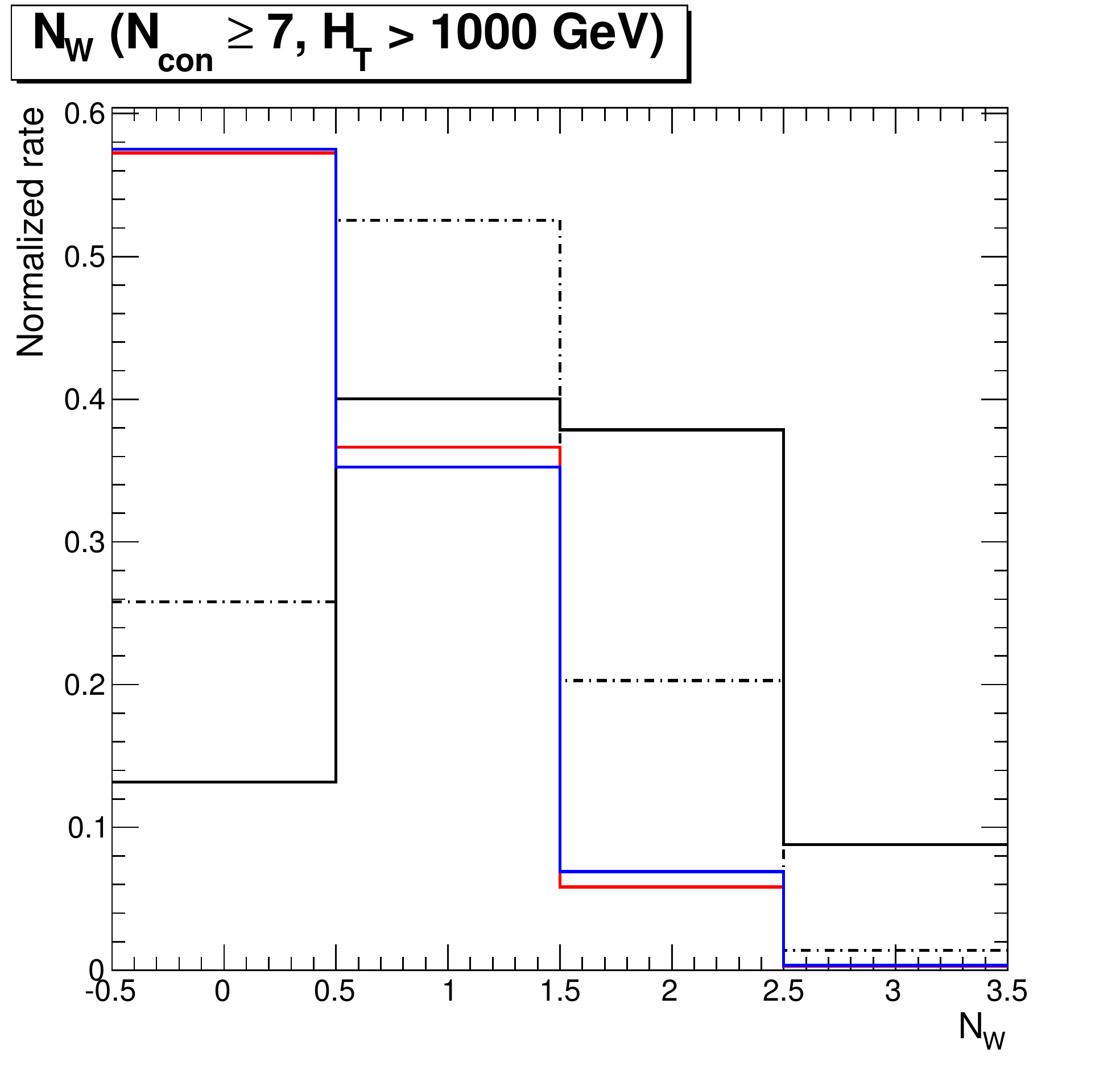}
\epsfxsize=0.31\textwidth\epsfbox{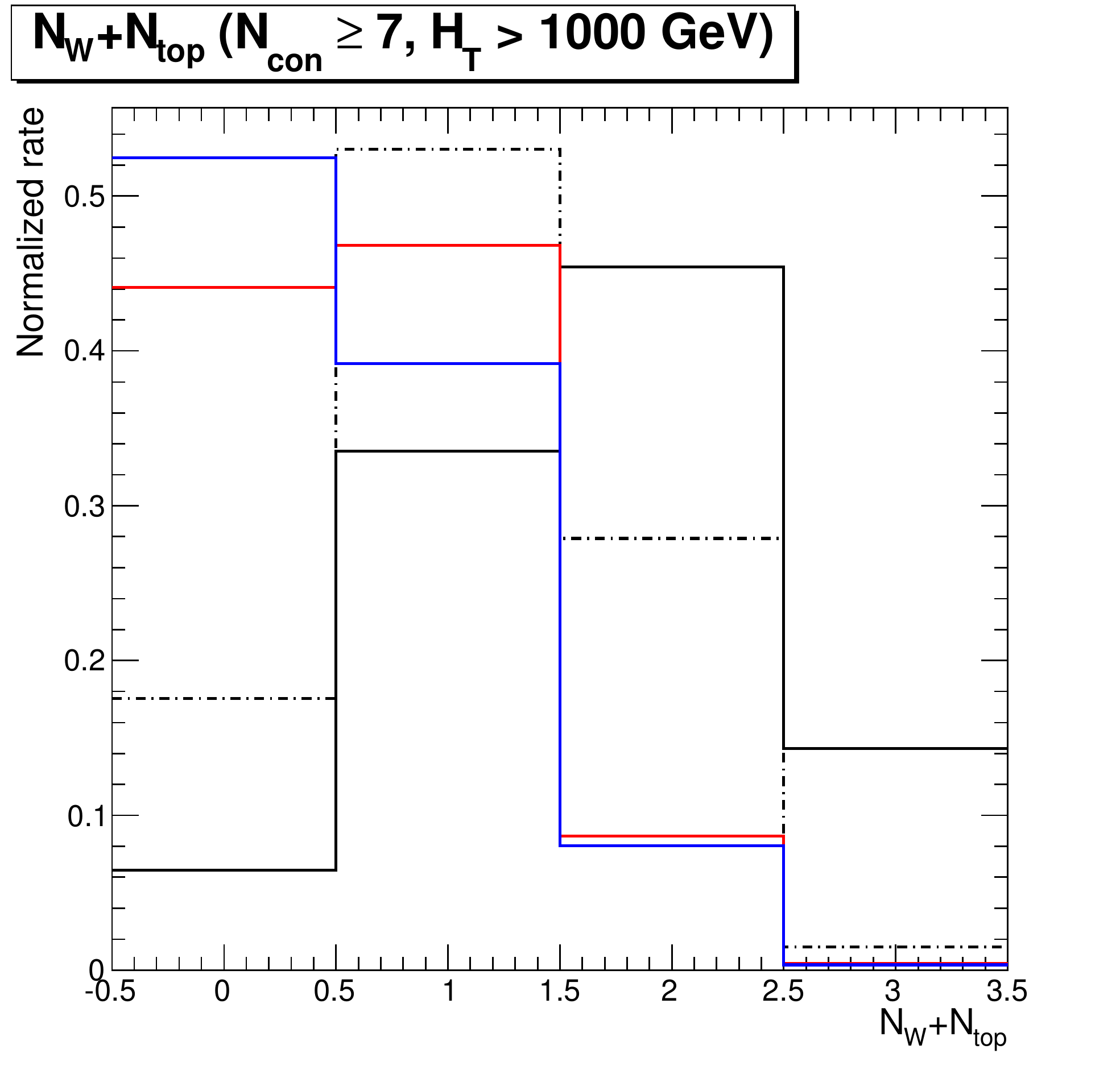}
\caption{Normalized distributions of the boosted tops (left), boosted $W$s (middle), and boosted tops + $W$s (right) in the $l$ + jets channel. Signals are pair (black solid) and single (black dashed) production of 800 GeV $X_{5/3}$, and backgrounds are $t\bar t$ + jets (solid red) and $W$ + jets (solid blue). Events are selected requiring $N_{\rm con} \geq 7$ and $H_T > 1000$ GeV.}
\label{fig:NboostOneLepton}
\end{center}
\end{figure}

The events are triggered by one isolated lepton. A lepton is considered isolated if the surrounding hadronic activity within a cone of size $R = 0.3$ satisfies $p_T(l)/(p_T(l)+p_T({\rm cone})) > 0.85$. The hadronic activity is organised such that objects with more substructure are looked at first, followed by object with less substructure in the $ttW +$ anything system. We cluster the event into $R=0.8$ ``fat-jets'' using the Cambridge/Aachen algorithm~\cite{Dokshitzer:1997in,Wobisch:1998wt}. We apply the HEPTopTagger~\cite{Plehn:2009rk, Plehn:2010st} on every fat-jet in the event to look for hadronic tops and remove them from the list of fat jets. Next we run the so-called BDRS tagger~\cite{Butterworth:2008iy} over the remaining fat-jets to look for hadronic $W$s. If the invariant mass of three filtered subjets meets the W mass requirement, $m^{\rm reco}_W = (65,95)$ GeV, we consider the jet to be a hadronic $W$-jet and remove it from the event.  All tagged hadronic top-jets or $W$-jets are required to be within $|\eta| < 2.5$. We collect all constituents from the remaining fat-jets that were not top or $W$ tagged, and we recluster them into $R=0.5$ anti-$k_T$~\cite{Cacciari:2008gp} jets. We accept only anti-$k_T$ jets with $p_T > 35$ GeV and $|\eta(j)| < 4.5$. Eventually we count the numbers of boosted tops and $W$s, and anti-$k_T$ jets and sum up the number of constituents $N_{\rm con}$ according to Eq.~\ref{eq:Ncon}.

We show in Fig.~\ref{fig:NjandNconOneLepton} the distributions of the traditional jet multiplicity, $N_j$, and the number of constituents, $N_{\rm con}$, for $X_{5/3}$ with 800 GeV and backgrounds respectively. The change of shapes between $N_\mathrm{con}$ and $N_j$ in the signal clearly indicates that using jet substructure methods is necessary to resolve the decay products of the top partners. In the background this effect is much less pronounced.

 In Fig.~\ref{fig:HTandNjOneLepton} we compare the $N_{\rm con}$ distributions of $X_{5/3}$ with 800 and 1000 GeV. The distributions are rather insensitive to the top partner masses, and they tend to have at least six (sub)jets. While single production is expected to have fewer constituents than pair production in general, the two distributions when restricted to the high $H_T$ region become very similar. We find that imposing $N_{\rm con} \geq 7$ is very effective to achieve a good statistical significance. We choose the $H_T$ cut such that we keep $\sim$90\% of the pair produced events after $N_{\rm con} \geq 7$ is imposed. The same $H_T$ cut keeps only $\sim$40-60\% of the single production events for the top partner mass range of interest. However, its smaller efficiency is compensated by the higher initial cross section in the single production process for heavy top partners. 

\begin{figure}
\begin{center}
\epsfxsize=0.46\textwidth\epsfbox{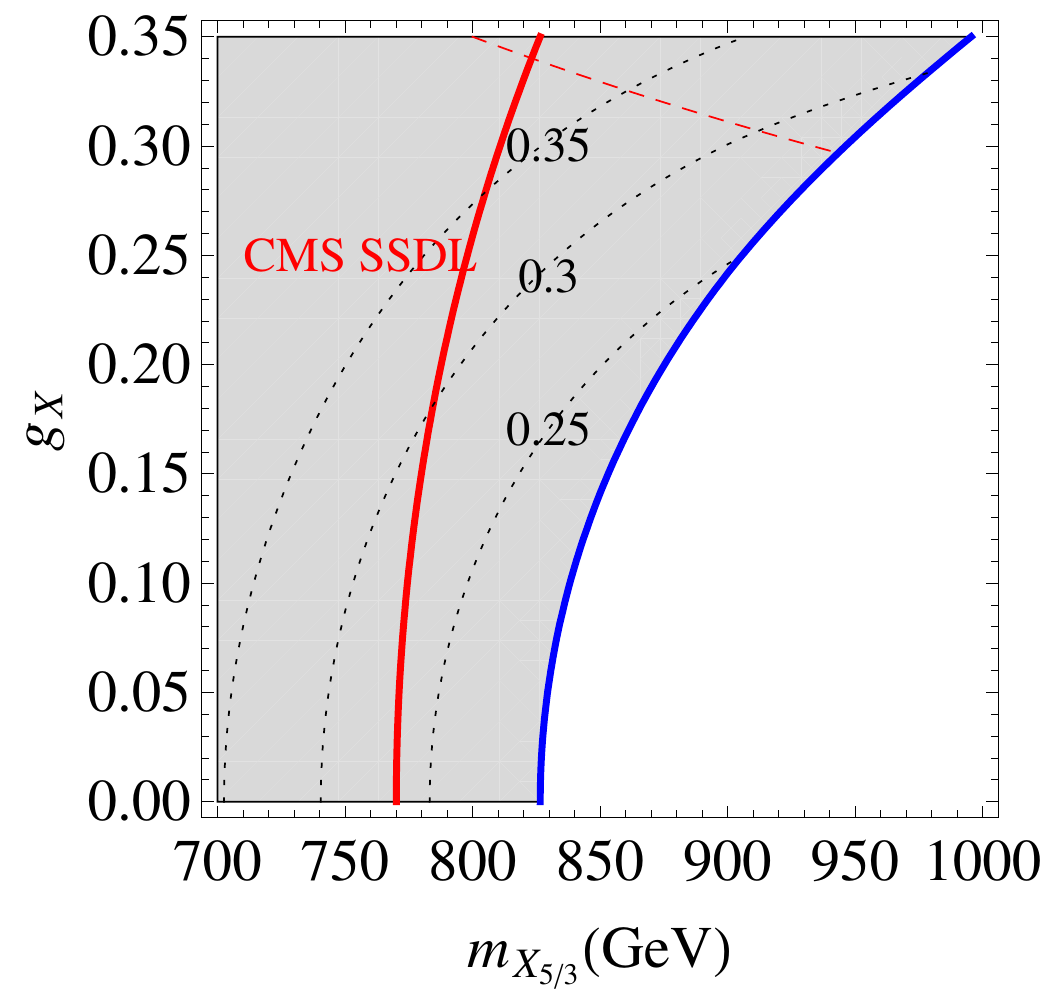}\quad
\epsfxsize=0.46\textwidth\epsfbox{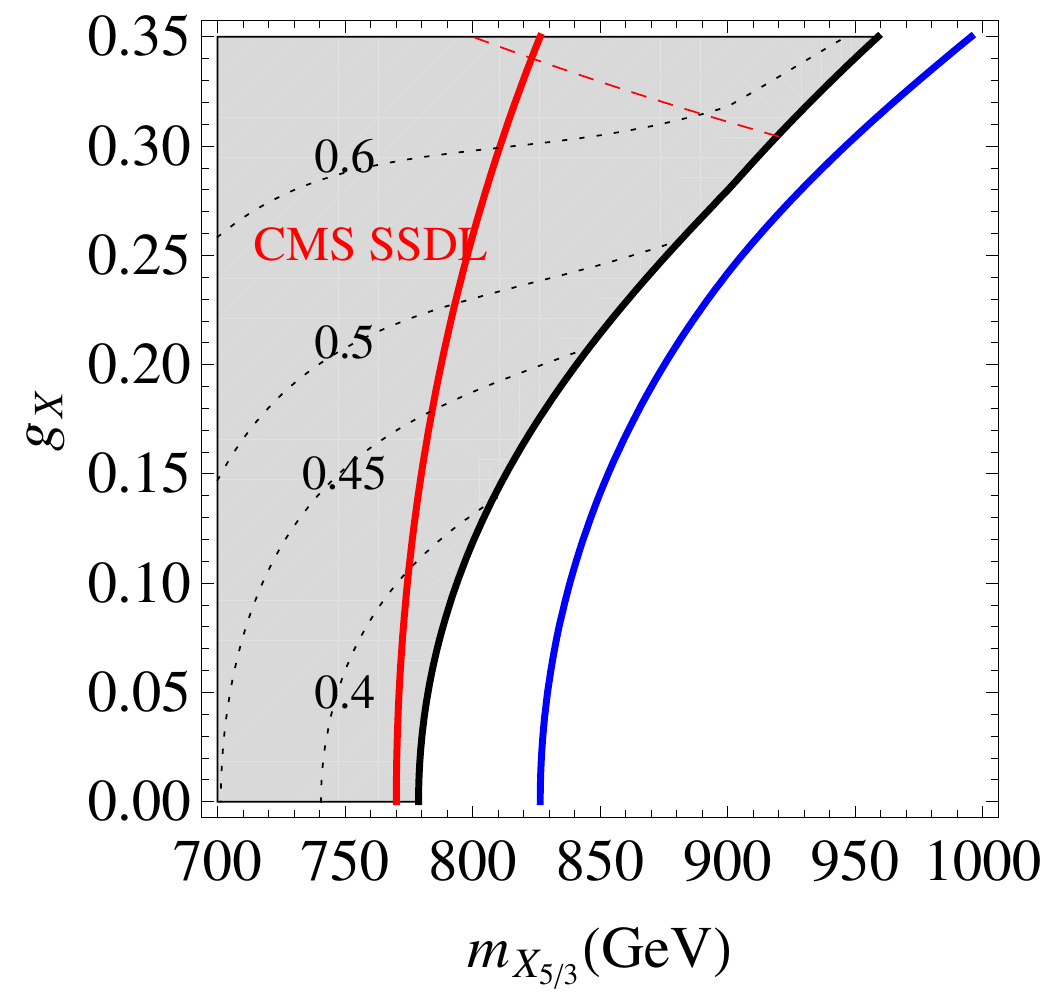}
\caption{Left: the estimated exclusion plot from our ``$l$ + jets'' cut-and-count analysis, assuming 20 fb$^{-1}$ of LHC8 data. Right: the corresponding exclusion plot after top partner mass reconstruction. The boundaries are defined by $S/\sqrt{S+B} = 2$. The solid red line displays the limit from the CMS SSDL search. The solid blue line in the right panel represents the exclusion curve from our cut-and-count analysis in the left panel. Black dotted lines indicate $S/B$ ratios. Red dashed lines indicate ${\mathcal O}(\phi^+ \bar X t) \sim g_X\,(m_{X}/m_W) = 3.5$.}
\label{fig:gVSmassLjets}
\end{center}
\end{figure}

We present the distributions of the boosted $W$-jets and top-jets, tagged using jet substructure techniques, in Fig.~\ref{fig:NboostOneLepton}. Imposing cuts on the total number of $W$-jets and top-jets is more effective than cutting on $N_W$ or $N_{\rm top}$ individually. The variable $N_{W}+N_{\rm top}$ is insensitive to whether a top is tagged as top-jet or only the $W$-boson of the top is tagged as $W$-jet. Thus we are less sensitive to the choice of the fat-jet's cone size\footnote{We explicitly studied the effect of the jet radius on $N_{W}+N_{\rm top}$ and find similar results.}. We demand $N_{W}+N_{\rm top} \geq 2$ which suppresses the backgrounds while keeping a handful of signal events of both single and pair production processes. One might want to try $N_{W}+N_{\rm top} \geq 3$, as suggested by the rightmost plot in Fig.~\ref{fig:NboostOneLepton}. While this more aggressive cut can significantly improve $S/B$, we lose the sensitivity to the single production process. We will not pursue this option, though we point out that at 14 TeV center-of-mass energy, with increased signal cross sections, this cut may improve the sensitivity of the search. After applying the outlined cuts on both pair and single production events as well as the backgrounds, we sum the expected signal events according to Eq.~\ref{eq:combine}. The estimated exclusion plot we show on the left panel of Fig.~\ref{fig:gVSmassLjets}.

%--------------------------------------
\subsection{Top partner mass reconstruction}
\label{subsec:toppartenrmassreco}
%--------------------------------------
\begin{figure}
\begin{center}
\epsfxsize=0.48\textwidth\epsfbox{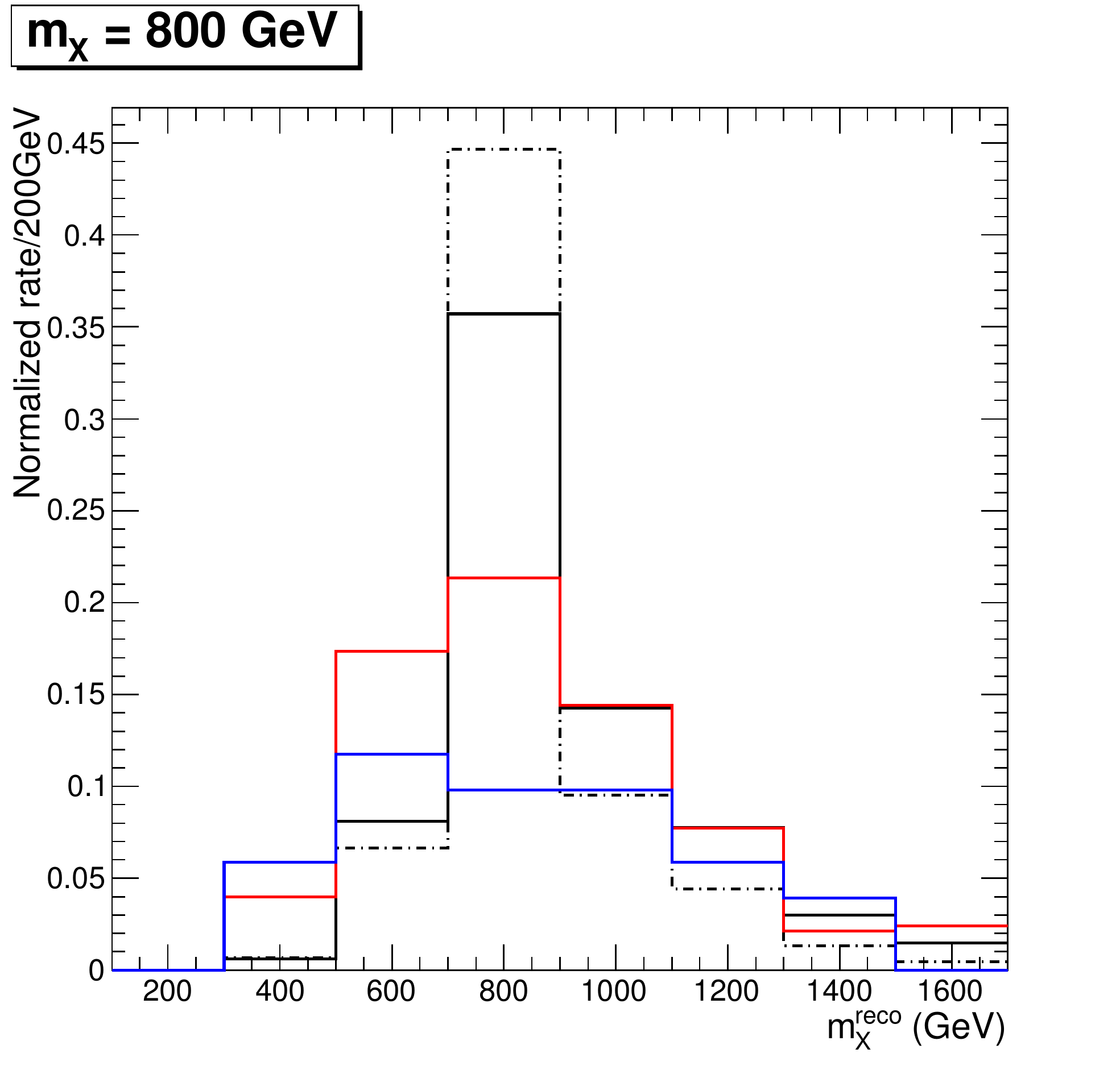}
\epsfxsize=0.48\textwidth\epsfbox{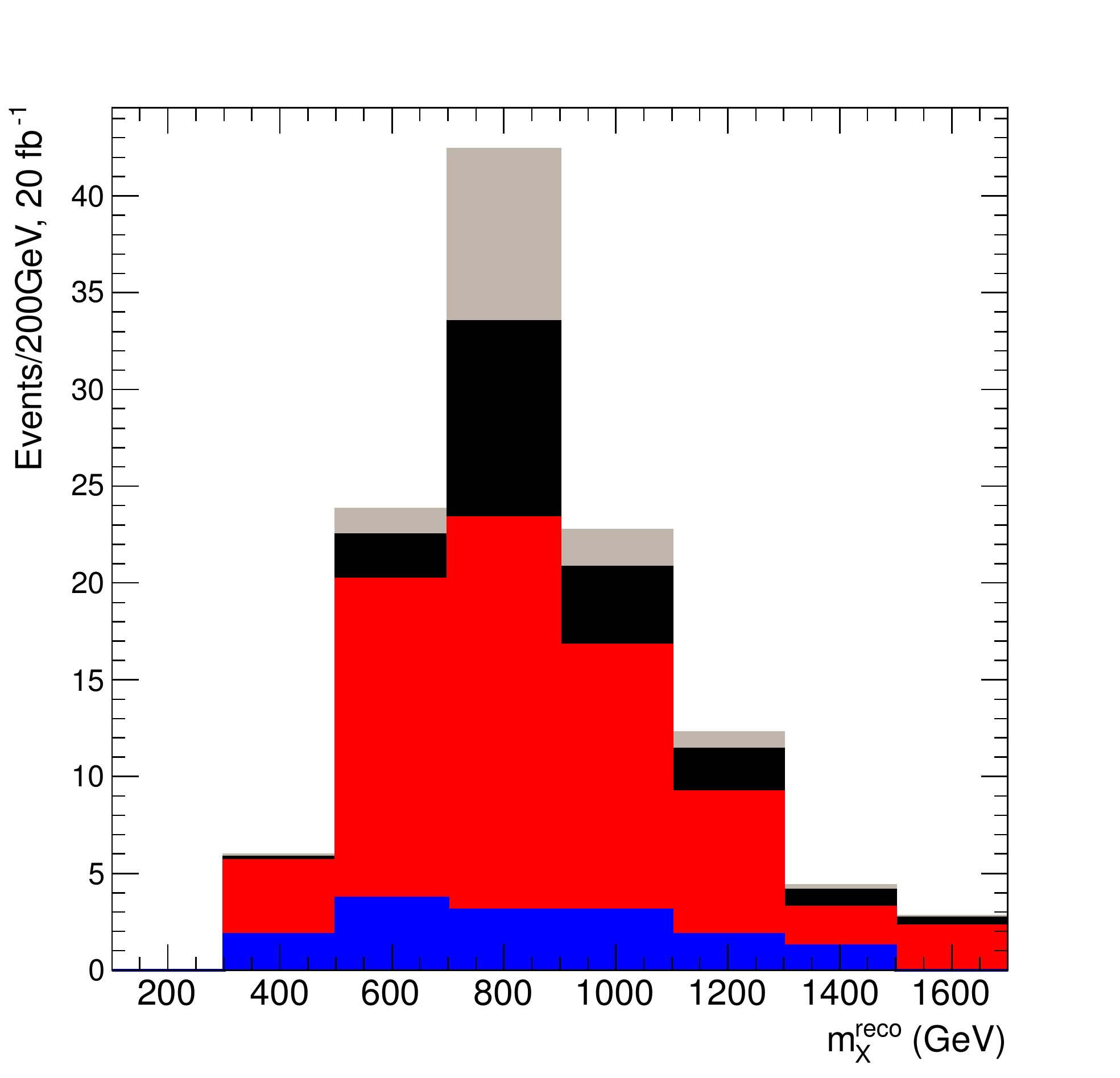}
\caption{The invariant mass of the reconstructed top partner. Left: normalized distributions of the signal's pair (solid black) and single (dashed black) production processes, and backgrounds, $t\bar t$ + jets (solid red) and $W$ + jets (solid blue). Right: stacked distributions of the signal's pair (black) and single (grey) production processes, and backgrounds, $t\bar t$ + jets (red) and $W$ + jets (blue), assuming 20 fb$^{-1}$ of LHC8 data. For the single production process we choose $g_X = 0.32$. All events are required to satisfy $N_{\rm con} \geq 7$, $H_T > 1000$ GeV and $N_W+N_{\rm top} \geq 2$.}
\label{fig:Xrecoljets}
\end{center}
\end{figure}

$X_{5/3}$ and $B$ decay both into a boosted top and W boson. Sometimes the top quark is not boosted enough to capture all decay products in a single jet and the b-jet and $W$ are reconstructed separately. Hence we iteratively pair each hadronic $W$-jet with any of the anti-$k_T$ jets within $\Delta R <$ 1.5 to look for a top candidate. If the invariant mass of the pair falls into the top mass window $m^{\rm reco}_{\rm top} = (150,\, 200)$ GeV and it satisfies $p_T (W j) > 200$ GeV and $|\eta(Wj)| < 2.5$, we count the jet pair as a top candidate. While our cut-and-count analysis does not care where the isolated lepton comes from, the reconstruction of the resonant top partner mass depends on its origin. Leptons are either produced from a $W$ originated in a top partner decay or a $W$ from an associated top quark in case of the single production process. In order to cover the maximum number of possibilities we reconstruct the leptonic $W$ and top quark as well. The leptonic $W$ is reconstructed by requiring $M_{\nu l}^2 = M_W^2$. We resolve the twofold ambiguity in calculating the longitudinal component of the neutrino momentum by choosing the solution which reconstructs the top mass best. After reconstructing the leptonic W we treat them on equal footing with the hadronic W: the leptonic $W$ is paired with any of the remaining anti-$k_T$ jets within $\Delta R <$ 1.5, and we consider the pair a leptonic top if it satisfies the same $p_T$ and $\eta$ requirements as the hadronic top, as well as $m_{lj} < 160$ GeV. Once an event is organised in terms of tops, $W$s, and everything else, we choose the top-$W$ pair with largest distance in azimuthal angle $\Delta \phi_{tW}$. The resulting reconstructed invariant masses of the top partners are shown in Fig.~\ref{fig:Xrecoljets}. The grey-coloured region in the right plot is contributed by the single production process for the coupling constant $g_X \sim 0.32$, which roughly translates to ${\mathcal O}(\phi^+ \bar X t) \sim 3.2$. We then count the number of signal and background events in the top partner mass window $m^{\rm reco}_{X} = (m_X-20\%,m_X+20\%)$ to estimate the sensitivity, assuming 20 fb$^{-1}$ of LHC8 data. 

\begin{table}[tbp]
\centering
{\scriptsize
\begin{tabular}{|c|c|c|c|c|c|c|}  
\hline
                   & $m_{X}$    & 700 GeV & 800 GeV  & 900 GeV & 1000 GeV & 1100 GeV \\ 
\hline \hline
                   & $H_T$ cut    & 900 \gev & 1000 \gev & 1110 \gev & 1200 \gev & 1300 \gev  \\ 
\hline
simple $(N_{\rm con},\ H_T)$  & pair production   & 6.4 fb  & 2.37 fb  & 0.93 fb  & 0.38 fb  & 0.16 fb  \\ \cline{2-7}
 $N_{\rm con} \ge 7$   & single - ($g_X=1$)    & 55.7 fb  & 33.9 fb  & 20.1 fb  & 12.7 fb  & 7.5 fb    \\ \cline{2-7}
    & $t\bar t$+jets  & 84.9 fb  & 52.2 fb  & 31 fb  & 20.1 fb  & 12.2 fb    \\ \cline{2-7}
    & $W$+jets        & 26.1 fb  & 18.9 fb  & 13.5 fb  & 10.2 fb  & 7.1 fb  \\ \cline{2-7}
\hline \hline
simple  $(N_{\rm con},H_T)$  & pair production  & 3.3 fb  & 1.4 fb  & 0.61 fb  & 0.26 fb  &  0.11 fb  \\ \cline{2-7}            
    $N_{\rm con} \ge 7$  & single - ($g_X=1$)    & 14.9 fb  & 10 fb  & 6.9 fb  & 4.6 fb  & 3 fb  \\ \cline{2-7}            
 + $(N_W+N_{\rm top}) \geq 2$  & $t\bar t$+jets & 7.1 fb  & 4.8 fb  & 3.2 fb  & 2 fb  & 1.3 fb  \\ \cline{2-7}
   & $W$+jets & 2.1 fb  & 1.6 fb  & 0.93 fb  & 0.62 fb  & 0.4 fb \\ \cline{2-7}
\hline \hline
$(N_j,H_T)$, $N_{\rm con} \ge 7$     & pair production    & 1.32 fb  & 0.64 fb  & 0.3 fb  & 0.13 fb  & 0.06 fb   \\ \cline{2-7}
  + $(N_W+N_{\rm top}) \geq 2$    & single - ($g_X=1$)      & 6.5 fb  & 5.1 fb  & 3.6 fb  & 2.6 fb  & 1.73 fb \\ \cline{2-7}
 + top partner reconstruction   & $t\bar t$+jets & 2.4 fb  & 1.5 fb  & 1 fb  & 0.75 fb  & 0.48 fb  \\ \cline{2-7}
with $0.8m_{X} < m^{\rm reco}_{X} < 1.2m_{X}$  & $W$+jets  & 0.52 fb  & 0.31 fb  & 0.16 fb  & 0.1 fb  & 0.06 fb \\ \cline{2-7}
\hline
\end{tabular}
}
\caption{Cross sections at LHC8 for the signals with different top partner masses and the corresponding backgrounds, after the different analysis cuts, in the $l$ + jets channel. $g_X$ is the coupling constant, involved in the production of the single top partner, and the numbers are for unit coupling.}
\label{tab:toppartner}
\end{table}

\begin{figure}
\begin{center}
\epsfxsize=0.50\textwidth\epsfbox{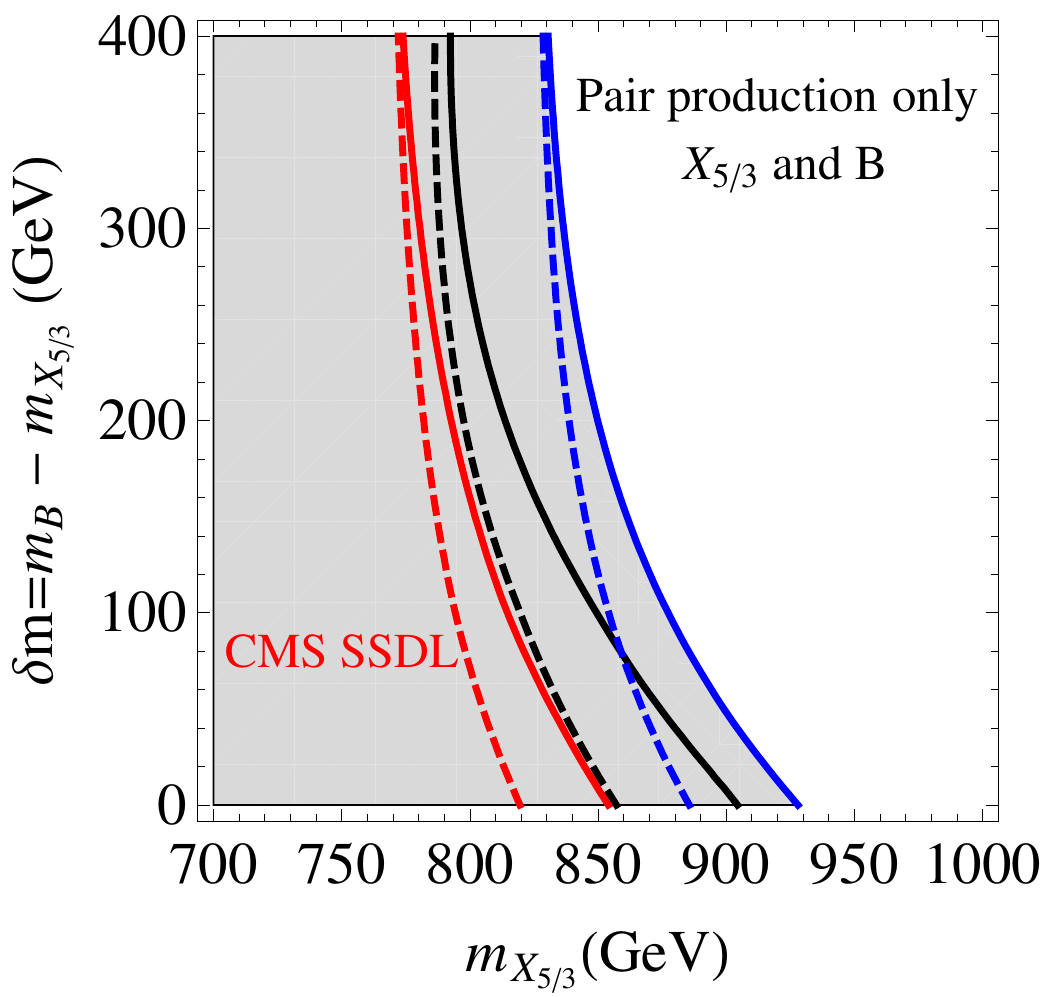}
\caption{The exclusion plot when the two contributions from $X_{5/3}$ and $B$ are summed in the total signal rate. The plot is restricted to only pair production processes of $X_{5/3}$ and $B$ for simplicity. BR($X_{5/3}\rightarrow tW$)=1 is assumed, and BR($B\rightarrow tW$)=0.5 (dashed) and 1 (solid) are plotted. Blue lines are obtained by our ``$l$ + jets'' style cut-and-count analysis, assuming 20 fb$^{-1}$ of LHC8 data. Black lines are after the top partner mass window is applied. Red lines indicates the recasted CMS SSDL.}
\label{fig:dmVSmassLjets}
\end{center}
\end{figure}

%\section{Implications on top partner models}
%\label{sec:application}
%\input{application}

\section{Discussion}
\label{sec:conclusions}

%%%%%%%%%%%%%%%%%%%%%%
%  Conclusions and Outlook
%%%%%%%%%%%%%%%%%%%%%%

\begin{figure}
\begin{center}
\epsfxsize=0.48\textwidth\epsfbox{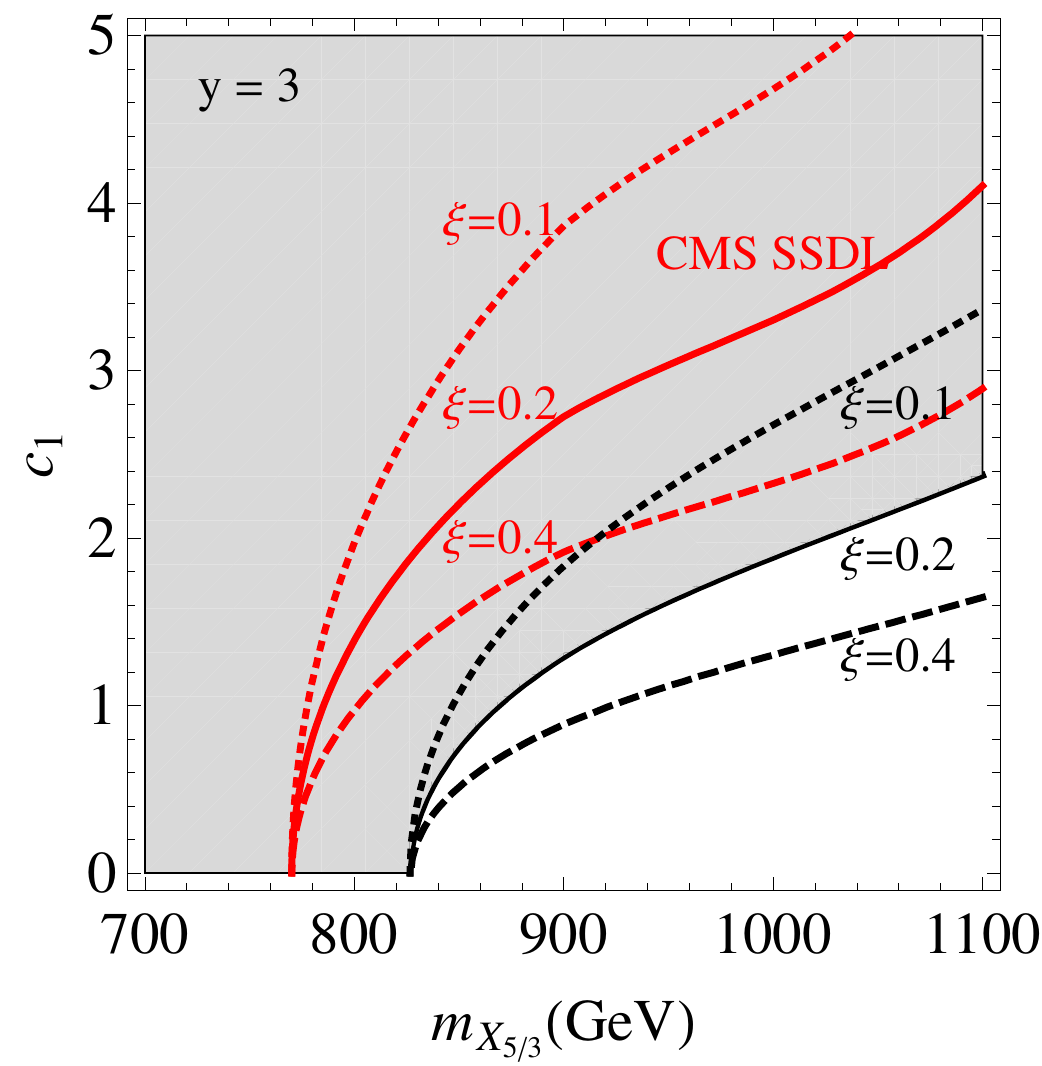}
\epsfxsize=0.48\textwidth\epsfbox{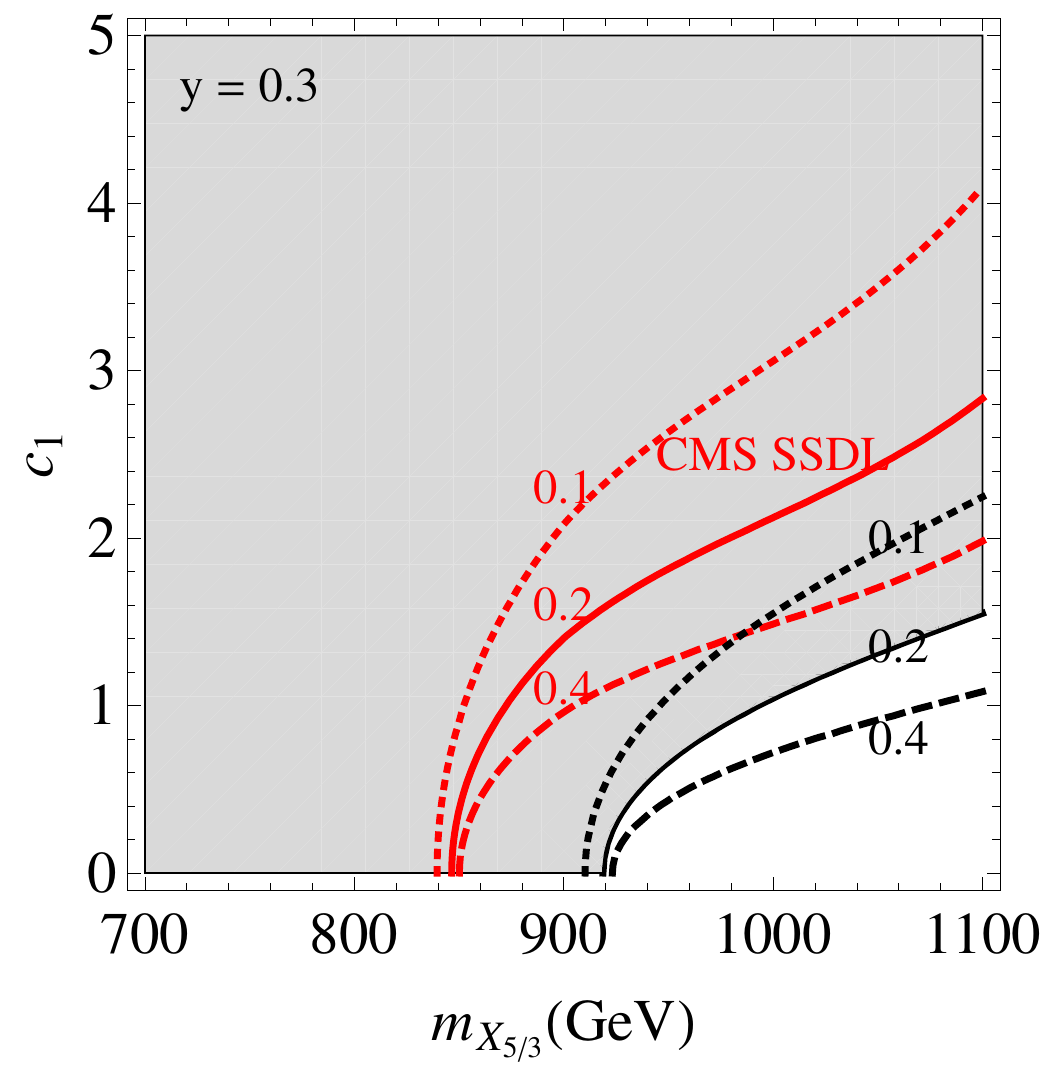}
\caption{The excluded region of ($c_1$, $m_X$) for a fixed set of $\xi$ and $y$. Two choices of $y$-values are shown. Left: $y=3$, corresponding to the case with $m_B \gg m_{5/3}$. Right: $y=0.3$, corresponding to the case with $m_B \gsim m_{5/3}$. For each $y$-value, we plot the contours for three different values of $\xi \equiv (v/f)^2$: $\xi =0.1$ (dotted), $\xi =0.2$ (solid), and $\xi =0.4$ (dashed). Black lines are obtained by our ``$l$ + jets'' style cut-and-count analysis, assuming 20 fb$^{-1}$ of LHC8 data. Red lines indicate the recast CMS SSDL analysis.}
\label{fig:c1vsMassOneLepton}
\end{center}
\end{figure}

Our results are summarised in a series of plots in Figs.~\ref{fig:gVSmassLjets},~\ref{fig:dmVSmassLjets}. We demonstrate that our one lepton analysis, taking into account the boosted kinematics, significantly improves the existing excluded region. Figs.~\ref{fig:gVSmassLjets} and~\ref{fig:dmVSmassLjets} show the relevant roles of two main ingredients, ``the correlation of processes'' and ``correlation of particles'', respectively, in improving the limit setting. For the former, in case the coupling $g_X$ of the single production process is large, the limit on the top partner $X_{5/3}$ can be increased up to $\sim 1$ TeV by our ``$l$ + jets'' style cut-and-count analysis. Exploiting the top partner mass reconstruction has little effect on the statistical significance. However, $S/B$ is improved by roughly a factor 2. For the latter, when the two masses of $X_{5/3}$ and $B$ are degenerate and their branching ratios to $tW$ are 1, their mass limit can be improved to $\sim 930$ GeV using the pair production process alone.

While Figs.~\ref{fig:gVSmassLjets} and \ref{fig:dmVSmassLjets} show the effect on the limits for the top partner masses, their implication for the composite Higgs model's input parameters is not transparent. Therefore we will rephrase our findings according to Eqs.~\ref{eq:Bmass},~\ref{eq:topmass} and \ref{eq:singleVertexM54} in terms of the free parameters of the theory. For instance we show in Fig.~\ref{fig:c1vsMassOneLepton} the excluded region of $c_1$ for varying top partner masses, while keeping $y$ and $\xi$ fixed. For further illustration we choose two values of $y$, representing two different mass hierarchies between $B$ and $X_{5/3}$ (see Eq.~\ref{eq:Bmass}). If $y = 3$, the contribution from $B$ is negligible compared to $X_{5/3}$ due to $m_B \gg m_{X_{5/3}}$, and the limit is mainly set by $X_{5/3}$ alone. For small $y$ (e.g. $y=0.3$ as in Fig.~\ref{fig:c1vsMassOneLepton}), the additional contribution from $B$ becomes very important. The way we defined the simplified model in Eq.~\ref{eq:M54}, the bottom-like top partner $B$ dominantly decays to $tW$ as decays to $bZ$ and $bh$ are forbidden~\cite{DeSimone:2012fs}. While the limit from direct searches for $B$ is identical to the limit on $X_{5/3}$, the indirect bound set by Eq.~\ref{eq:Bmass} is higher. For $y = 3$ and $\xi = 0.1-0.4$ in Fig.~\ref{fig:c1vsMassOneLepton}, the bounds on $m_{5/3}$ indicate that $B$ can be excluded at masses below $m_B \gsim 1.5 - 2.5$ TeV. For the squeezed spectrum with $y=0.3$,  as shown in Fig.~\ref{fig:c1vsMassOneLepton}, the $B$ mass is expected to be larger than $m_B \gsim 930 - 940$ GeV assuming $\xi = 0.1-0.4$. The mass splitting between $B$ and $T$ after EWSB is much smaller than the splitting between the SU(2) doublets ($B$, $T$) and ($X_{5/3}$, $X_{2/3}$) (see Eq~\ref{eq:topmass}). The indirect bound on $T$, derived from the direct limit on $X_{5/3}$, is very similar to the one of $B$.\footnote{The parameters in our simplified model are also indirectly constrained by electroweak precision measurements (EWPM) \cite{Grojean:2013qca}.}

While we show exclusion bounds in Figs.~\ref{fig:gVSmassLjets} and \ref{fig:dmVSmassLjets} only for a few benchmark values of the parameters, from the set of contours one can extract other choices easily. The results shown in Figs.~\ref{fig:gVSmassLjets} and \ref{fig:dmVSmassLjets} are not restricted to the simplified model as defined in Eq.~\ref{eq:M54} only. Their benefit extends to any heavy vector-like fermion model that shares the same decay topology. Therefore, the parameters of those models can be constrained by our results, shown in Figs.~\ref{fig:gVSmassLjets} and~\ref{fig:dmVSmassLjets}.\\

So far in this work we only discussed results for 8 TeV center-of-mass energy. At the end of 2014 the LHC is going to restart with $\sqrt{s}=13-14~\rm{TeV}$. At this energy, due to the lower Bjorken-$x$ needed to produce the top partners, the production cross section for the gluon-induced pair production process will be strongly increased. Thus, the cross over point where the single production process has the same production cross section as the pair production process will be shifted to larger top partner masses. Still, our finding that exploiting the correlation of different production processes and contributions from different top partners to the $t\bar{t}W$ final state is beneficial in constraining the free parameters of the composite Higgs model carries over straightforwardly. 
As heavier top partners can be probed at 14 TeV their decay products will be more boosted and their radiation will be confined to a smaller area of the detector. Particularly for the reconstruction of isolated leptons this can pose a severe challenge. However, already in searches at 8 TeV mini-isolation criteria for the reconstruction of isolated leptons were proposed and successfully applied \cite{Rehermann:2010vq}. In this kinematic regime boosted techniques will be indispensable. In fact, some of the existing taggers might need further development to exploit the LHC's energy reach to the fullest \cite{Schaetzel:2013vka}. In any case, the observables and search strategies discussed in this work will be directly applicable at 13 (14) TeV, hereby helping to discover TeV-scale top partners or constraining the parameter space of composite Higgs models.

\appendix

%\section{Simulation}
%\label{app:simulations}
%\input{simulations}

%%%%%%%%%%%%%%%%%%%%%%%%%%%%%%%%%%%%%%%%%%%%%%%%%%%%%%%%%%%%%%%%%%%%%%%%

\acknowledgments{We thank Zhenyu Han and Brock Tweedie for collaboration at an early stage of this work. We thank Christopher Wallace for many helpful comments on our draft. We thank Giacomo Cacciapaglia, Roberto Contino, Zhenyu Han, Tobias Golling, Seung Lee, Brock Tweedie, Andrea Wulzer for valuable discussions. Two MS thank the organisers of the workshop {\it Physics at TeV Colliders} at Les Houches for hospitality. AA and MS were supported by the ERC Advanced Grant No. 267985, ``Electroweak Symmetry Breaking, Flavour and Dark Matter: One Solution for Three Mysteries'' (DaMeSyFla).  }

%%%%%%%%%%%%%%%%%%%%%%%%%%%%%%%%%%%%%%%%%%%%%%%%%%%%%%%%%%%%%%%%%%%%%%%

%%%%%%%%%%%%%%
% References
%%%%%%%%%%%%%%

\bibliography{lit}

\begin{thebibliography}{65}
\expandafter\ifx\csname natexlab\endcsname\relax\def\natexlab#1{#1}\fi
\expandafter\ifx\csname bibnamefont\endcsname\relax
  \def\bibnamefont#1{#1}\fi
\expandafter\ifx\csname bibfnamefont\endcsname\relax
  \def\bibfnamefont#1{#1}\fi
\expandafter\ifx\csname citenamefont\endcsname\relax
  \def\citenamefont#1{#1}\fi
\expandafter\ifx\csname url\endcsname\relax
  \def\url#1{\texttt{#1}}\fi
\expandafter\ifx\csname urlprefix\endcsname\relax\def\urlprefix{URL }\fi
\providecommand{\bibinfo}[2]{#2}
\providecommand{\eprint}[2][]{\url{#2}}

\bibitem[{\citenamefont{Aad et~al.}(2012)}]{Aad:2012tfa}
\bibinfo{author}{\bibfnamefont{G.}~\bibnamefont{Aad}} \bibnamefont{et~al.}
  (\bibinfo{collaboration}{ATLAS Collaboration}),
  \emph{\bibinfo{title}{{Observation of a new particle in the search for the
  Standard Model Higgs boson with the ATLAS detector at the LHC}}},
  \bibinfo{journal}{Phys.Lett.} \textbf{\bibinfo{volume}{B716}},
  \bibinfo{pages}{1} (\bibinfo{year}{2012}), \eprint{1207.7214}.

\bibitem[{\citenamefont{Chatrchyan et~al.}(2012)}]{Chatrchyan:2012ufa}
\bibinfo{author}{\bibfnamefont{S.}~\bibnamefont{Chatrchyan}}
  \bibnamefont{et~al.} (\bibinfo{collaboration}{CMS Collaboration}),
  \emph{\bibinfo{title}{{Observation of a new boson at a mass of 125 GeV with
  the CMS experiment at the LHC}}}, \bibinfo{journal}{Phys.Lett.}
  \textbf{\bibinfo{volume}{B716}}, \bibinfo{pages}{30} (\bibinfo{year}{2012}),
  \eprint{1207.7235}.

\bibitem[{\citenamefont{Kaplan and Georgi}(1984)}]{Kaplan:1983fs}
\bibinfo{author}{\bibfnamefont{D.~B.} \bibnamefont{Kaplan}} \bibnamefont{and}
  \bibinfo{author}{\bibfnamefont{H.}~\bibnamefont{Georgi}},
  \emph{\bibinfo{title}{{SU(2) x U(1) Breaking by Vacuum Misalignment}}},
  \bibinfo{journal}{Phys. Lett.} \textbf{\bibinfo{volume}{B136}},
  \bibinfo{pages}{183} (\bibinfo{year}{1984}).

\bibitem[{\citenamefont{Kaplan et~al.}(1984)\citenamefont{Kaplan, Georgi, and
  Dimopoulos}}]{Kaplan:1983sm}
\bibinfo{author}{\bibfnamefont{D.~B.} \bibnamefont{Kaplan}},
  \bibinfo{author}{\bibfnamefont{H.}~\bibnamefont{Georgi}}, \bibnamefont{and}
  \bibinfo{author}{\bibfnamefont{S.}~\bibnamefont{Dimopoulos}},
  \emph{\bibinfo{title}{{Composite Higgs Scalars}}}, \bibinfo{journal}{Phys.
  Lett.} \textbf{\bibinfo{volume}{B136}}, \bibinfo{pages}{187}
  (\bibinfo{year}{1984}).

\bibitem[{\citenamefont{Georgi and Kaplan}(1984)}]{Georgi:1984af}
\bibinfo{author}{\bibfnamefont{H.}~\bibnamefont{Georgi}} \bibnamefont{and}
  \bibinfo{author}{\bibfnamefont{D.~B.} \bibnamefont{Kaplan}},
  \emph{\bibinfo{title}{{Composite Higgs and Custodial SU(2)}}},
  \bibinfo{journal}{Phys. Lett.} \textbf{\bibinfo{volume}{B145}},
  \bibinfo{pages}{216} (\bibinfo{year}{1984}).

\bibitem[{\citenamefont{Dugan et~al.}(1985)\citenamefont{Dugan, Georgi, and
  Kaplan}}]{Dugan:1984hq}
\bibinfo{author}{\bibfnamefont{M.~J.} \bibnamefont{Dugan}},
  \bibinfo{author}{\bibfnamefont{H.}~\bibnamefont{Georgi}}, \bibnamefont{and}
  \bibinfo{author}{\bibfnamefont{D.~B.} \bibnamefont{Kaplan}},
  \emph{\bibinfo{title}{{Anatomy of a Composite Higgs Model}}},
  \bibinfo{journal}{Nucl. Phys.} \textbf{\bibinfo{volume}{B254}},
  \bibinfo{pages}{299} (\bibinfo{year}{1985}).

\bibitem[{\citenamefont{Contino et~al.}(2003)\citenamefont{Contino, Nomura, and
  Pomarol}}]{Contino:2003ve}
\bibinfo{author}{\bibfnamefont{R.}~\bibnamefont{Contino}},
  \bibinfo{author}{\bibfnamefont{Y.}~\bibnamefont{Nomura}}, \bibnamefont{and}
  \bibinfo{author}{\bibfnamefont{A.}~\bibnamefont{Pomarol}},
  \emph{\bibinfo{title}{{Higgs as a Holographic Pseudo-Goldstone Boson}}},
  \bibinfo{journal}{Nucl. Phys.} \textbf{\bibinfo{volume}{B671}},
  \bibinfo{pages}{148} (\bibinfo{year}{2003}), \eprint{hep-ph/0306259}.

\bibitem[{\citenamefont{Contino}(2010)}]{Contino:2010rs}
\bibinfo{author}{\bibfnamefont{R.}~\bibnamefont{Contino}},
  \emph{\bibinfo{title}{{The Higgs as a Composite Nambu-Goldstone Boson}}}
  (\bibinfo{year}{2010}), \eprint{1005.4269}.

\bibitem[{\citenamefont{Keren-Zur et~al.}(2012)\citenamefont{Keren-Zur, Lodone,
  Nardecchia, Pappadopulo, Rattazzi et~al.}}]{KerenZur:2012fr}
\bibinfo{author}{\bibfnamefont{B.}~\bibnamefont{Keren-Zur}},
  \bibinfo{author}{\bibfnamefont{P.}~\bibnamefont{Lodone}},
  \bibinfo{author}{\bibfnamefont{M.}~\bibnamefont{Nardecchia}},
  \bibinfo{author}{\bibfnamefont{D.}~\bibnamefont{Pappadopulo}},
  \bibinfo{author}{\bibfnamefont{R.}~\bibnamefont{Rattazzi}},
  \bibnamefont{et~al.}, \emph{\bibinfo{title}{{On Partial Compositeness and the
  CP asymmetry in charm decays}}} (\bibinfo{year}{2012}), \eprint{1205.5803}.

\bibitem[{\citenamefont{Kaplan}(1991)}]{Kaplan:1991dc}
\bibinfo{author}{\bibfnamefont{D.~B.} \bibnamefont{Kaplan}},
  \emph{\bibinfo{title}{{Flavor at SSC energies: A New Mechanism for
  Dynamically Generated Fermion Masses}}}, \bibinfo{journal}{Nucl. Phys.}
  \textbf{\bibinfo{volume}{B365}}, \bibinfo{pages}{259} (\bibinfo{year}{1991}).

\bibitem[{\citenamefont{Agashe et~al.}(2006)\citenamefont{Agashe, Contino, {Da
  Rold}, and Pomarol}}]{Agashe:2006at}
\bibinfo{author}{\bibfnamefont{K.}~\bibnamefont{Agashe}},
  \bibinfo{author}{\bibfnamefont{R.}~\bibnamefont{Contino}},
  \bibinfo{author}{\bibfnamefont{L.}~\bibnamefont{{Da Rold}}},
  \bibnamefont{and} \bibinfo{author}{\bibfnamefont{A.}~\bibnamefont{Pomarol}},
  \emph{\bibinfo{title}{{A Custodial Symmetry for $Z b \bar b$}}},
  \bibinfo{journal}{Phys. Lett.} \textbf{\bibinfo{volume}{B641}},
  \bibinfo{pages}{62} (\bibinfo{year}{2006}), \eprint{hep-ph/0605341}.

\bibitem[{\citenamefont{Contino
  et~al.}(2007{\natexlab{a}})\citenamefont{Contino, Da~Rold, and
  Pomarol}}]{Contino:2006qr}
\bibinfo{author}{\bibfnamefont{R.}~\bibnamefont{Contino}},
  \bibinfo{author}{\bibfnamefont{L.}~\bibnamefont{Da~Rold}}, \bibnamefont{and}
  \bibinfo{author}{\bibfnamefont{A.}~\bibnamefont{Pomarol}},
  \emph{\bibinfo{title}{{Light custodians in natural composite Higgs models}}},
  \bibinfo{journal}{Phys.Rev.} \textbf{\bibinfo{volume}{D75}},
  \bibinfo{pages}{055014} (\bibinfo{year}{2007}{\natexlab{a}}),
  \eprint{hep-ph/0612048}.

\bibitem[{\citenamefont{Mrazek and Wulzer}(2010)}]{Mrazek:2009yu}
\bibinfo{author}{\bibfnamefont{J.}~\bibnamefont{Mrazek}} \bibnamefont{and}
  \bibinfo{author}{\bibfnamefont{A.}~\bibnamefont{Wulzer}},
  \emph{\bibinfo{title}{{A Strong Sector at the LHC: Top Partners in Same-Sign
  Dileptons}}}, \bibinfo{journal}{Phys.Rev.} \textbf{\bibinfo{volume}{D81}},
  \bibinfo{pages}{075006} (\bibinfo{year}{2010}), \eprint{0909.3977}.

\bibitem[{\citenamefont{Cacciapaglia et~al.}(2012)\citenamefont{Cacciapaglia,
  Deandrea, Panizzi, Gaur, Harada et~al.}}]{Cacciapaglia:2011fx}
\bibinfo{author}{\bibfnamefont{G.}~\bibnamefont{Cacciapaglia}},
  \bibinfo{author}{\bibfnamefont{A.}~\bibnamefont{Deandrea}},
  \bibinfo{author}{\bibfnamefont{L.}~\bibnamefont{Panizzi}},
  \bibinfo{author}{\bibfnamefont{N.}~\bibnamefont{Gaur}},
  \bibinfo{author}{\bibfnamefont{D.}~\bibnamefont{Harada}},
  \bibnamefont{et~al.}, \emph{\bibinfo{title}{{Heavy Vector-like Top Partners
  at the LHC and flavour constraints}}}, \bibinfo{journal}{JHEP}
  \textbf{\bibinfo{volume}{1203}}, \bibinfo{pages}{070} (\bibinfo{year}{2012}),
  \eprint{1108.6329}.

\bibitem[{\citenamefont{Vignaroli}(2012{\natexlab{a}})}]{Vignaroli:2012nf}
\bibinfo{author}{\bibfnamefont{N.}~\bibnamefont{Vignaroli}},
  \emph{\bibinfo{title}{{Early discovery of top partners and test of the Higgs
  nature}}}, \bibinfo{journal}{Phys.Rev.} \textbf{\bibinfo{volume}{D86}},
  \bibinfo{pages}{075017} (\bibinfo{year}{2012}{\natexlab{a}}),
  \eprint{1207.0830}.

\bibitem[{\citenamefont{Vignaroli}(2012{\natexlab{b}})}]{Vignaroli:2012sf}
\bibinfo{author}{\bibfnamefont{N.}~\bibnamefont{Vignaroli}},
  \emph{\bibinfo{title}{{Discovering the composite Higgs through the decay of a
  heavy fermion}}}, \bibinfo{journal}{JHEP} \textbf{\bibinfo{volume}{1207}},
  \bibinfo{pages}{158} (\bibinfo{year}{2012}{\natexlab{b}}),
  \eprint{1204.0468}.

\bibitem[{\citenamefont{Li et~al.}(2013)\citenamefont{Li, Liu, and
  Shu}}]{Li:2013xba}
\bibinfo{author}{\bibfnamefont{J.}~\bibnamefont{Li}},
  \bibinfo{author}{\bibfnamefont{D.}~\bibnamefont{Liu}}, \bibnamefont{and}
  \bibinfo{author}{\bibfnamefont{J.}~\bibnamefont{Shu}},
  \emph{\bibinfo{title}{{Towards the fate of natural composite Higgs model
  through single $t^\prime$ search at the 8 TeV LHC}}} (\bibinfo{year}{2013}),
  \eprint{1306.5841}.

\bibitem[{\citenamefont{De~Simone et~al.}(2013)\citenamefont{De~Simone,
  Matsedonskyi, Rattazzi, and Wulzer}}]{DeSimone:2012fs}
\bibinfo{author}{\bibfnamefont{A.}~\bibnamefont{De~Simone}},
  \bibinfo{author}{\bibfnamefont{O.}~\bibnamefont{Matsedonskyi}},
  \bibinfo{author}{\bibfnamefont{R.}~\bibnamefont{Rattazzi}}, \bibnamefont{and}
  \bibinfo{author}{\bibfnamefont{A.}~\bibnamefont{Wulzer}},
  \emph{\bibinfo{title}{{A First Top Partner's Hunter Guide}}},
  \bibinfo{journal}{JHEP} \textbf{\bibinfo{volume}{1304}}, \bibinfo{pages}{004}
  (\bibinfo{year}{2013}), \eprint{1211.5663}.

\bibitem[{ATL(2013{\natexlab{a}})}]{ATLAS-CONF-2013-051}
\bibinfo{type}{Tech. Rep.} \bibinfo{number}{ATLAS-CONF-2013-051},
  \bibinfo{institution}{CERN}, \bibinfo{address}{Geneva}
  (\bibinfo{year}{2013}{\natexlab{a}}).

\bibitem[{ATL(2013{\natexlab{b}})}]{ATLAS-CONF-2013-056}
\bibinfo{type}{Tech. Rep.} \bibinfo{number}{ATLAS-CONF-2013-056},
  \bibinfo{institution}{CERN}, \bibinfo{address}{Geneva}
  (\bibinfo{year}{2013}{\natexlab{b}}).

\bibitem[{ATL(2013{\natexlab{c}})}]{ATLAS-CONF-2013-060}
\bibinfo{type}{Tech. Rep.} \bibinfo{number}{ATLAS-CONF-2013-060},
  \bibinfo{institution}{CERN}, \bibinfo{address}{Geneva}
  (\bibinfo{year}{2013}{\natexlab{c}}).

\bibitem[{1229964()}]{ATLAS:2013ima}
1229964, \emph{\bibinfo{title}{{Search for heavy top-like quarks decaying to a
  Higgs boson and a top quark in the lepton plus jets final state in $pp$
  collisions at $\sqrt{s}=8$ TeV with the ATLAS detector}}}
  (\bibinfo{year}{2013}).

\bibitem[{CMS(2013{\natexlab{a}})}]{CMS:2013sin}
\emph{\bibinfo{title}{{Inclusive search for vector-like T-quark by CMS}}}
  (\bibinfo{year}{2013}{\natexlab{a}}).

\bibitem[{1230237()}]{CMS:vwa}
1230237, \emph{\bibinfo{title}{{Search for T5/3 top partners in same-sign
  dilepton final state}}} (\bibinfo{year}{2013}).

\bibitem[{CMS(2013{\natexlab{b}})}]{CMS-PAS-B2G-12-021}
\bibinfo{type}{Tech. Rep.} \bibinfo{number}{CMS-PAS-B2G-12-021},
  \bibinfo{institution}{CERN}, \bibinfo{address}{Geneva}
  (\bibinfo{year}{2013}{\natexlab{b}}).

\bibitem[{CMS(2012)}]{CMS-PAS-SUS-12-027}
\bibinfo{type}{Tech. Rep.} \bibinfo{number}{CMS-PAS-SUS-12-027},
  \bibinfo{institution}{CERN}, \bibinfo{address}{Geneva}
  (\bibinfo{year}{2012}).

\bibitem[{\citenamefont{Brooijmans}()}]{Brooijmans:2008zza}
\bibinfo{author}{\bibfnamefont{G.}~\bibnamefont{Brooijmans}},
  \emph{\bibinfo{title}{{High $p_T$ Hadronic Top Quark Identification. Part I:
  Jet Mass and YSplitter}}}, \bibinfo{note}{{ATL-PHYS-CONF-2008-008}}.

\bibitem[{\citenamefont{Thaler and Wang}(2008)}]{Thaler:2008ju}
\bibinfo{author}{\bibfnamefont{J.}~\bibnamefont{Thaler}} \bibnamefont{and}
  \bibinfo{author}{\bibfnamefont{L.-T.} \bibnamefont{Wang}},
  \emph{\bibinfo{title}{{Strategies to Identify Boosted Tops}}},
  \bibinfo{journal}{JHEP} \textbf{\bibinfo{volume}{07}}, \bibinfo{pages}{092}
  (\bibinfo{year}{2008}), \eprint{0806.0023}.

\bibitem[{\citenamefont{Kaplan et~al.}(2008)\citenamefont{Kaplan, Rehermann,
  Schwartz, and Tweedie}}]{Kaplan:2008ie}
\bibinfo{author}{\bibfnamefont{D.~E.} \bibnamefont{Kaplan}},
  \bibinfo{author}{\bibfnamefont{K.}~\bibnamefont{Rehermann}},
  \bibinfo{author}{\bibfnamefont{M.~D.} \bibnamefont{Schwartz}},
  \bibnamefont{and} \bibinfo{author}{\bibfnamefont{B.}~\bibnamefont{Tweedie}},
  \emph{\bibinfo{title}{{Top Tagging: A Method for Identifying Boosted
  Hadronically Decaying Top Quarks}}}, \bibinfo{journal}{Phys. Rev. Lett.}
  \textbf{\bibinfo{volume}{101}}, \bibinfo{pages}{142001}
  (\bibinfo{year}{2008}), \eprint{0806.0848}.

\bibitem[{\citenamefont{Almeida et~al.}(2009)}]{Almeida:2008yp}
\bibinfo{author}{\bibfnamefont{L.~G.} \bibnamefont{Almeida}}
  \bibnamefont{et~al.}, \emph{\bibinfo{title}{{Substructure of High-$p_T$ Jets
  at the LHC}}}, \bibinfo{journal}{Phys. Rev.} \textbf{\bibinfo{volume}{D79}},
  \bibinfo{pages}{074017} (\bibinfo{year}{2009}), \eprint{0807.0234}.

\bibitem[{\citenamefont{Almeida et~al.}(2010)\citenamefont{Almeida, Lee, Perez,
  Sterman, and Sung}}]{Almeida:2010pa}
\bibinfo{author}{\bibfnamefont{L.~G.} \bibnamefont{Almeida}},
  \bibinfo{author}{\bibfnamefont{S.~J.} \bibnamefont{Lee}},
  \bibinfo{author}{\bibfnamefont{G.}~\bibnamefont{Perez}},
  \bibinfo{author}{\bibfnamefont{G.}~\bibnamefont{Sterman}}, \bibnamefont{and}
  \bibinfo{author}{\bibfnamefont{I.}~\bibnamefont{Sung}},
  \emph{\bibinfo{title}{{Template Overlap Method for Massive Jets}}},
  \bibinfo{journal}{Phys. Rev.} \textbf{\bibinfo{volume}{D82}},
  \bibinfo{pages}{054034} (\bibinfo{year}{2010}), \eprint{1006.2035}.

\bibitem[{\citenamefont{Plehn et~al.}(2010{\natexlab{a}})\citenamefont{Plehn,
  Spannowsky, Takeuchi, and Zerwas}}]{Plehn:2010st}
\bibinfo{author}{\bibfnamefont{T.}~\bibnamefont{Plehn}},
  \bibinfo{author}{\bibfnamefont{M.}~\bibnamefont{Spannowsky}},
  \bibinfo{author}{\bibfnamefont{M.}~\bibnamefont{Takeuchi}}, \bibnamefont{and}
  \bibinfo{author}{\bibfnamefont{D.}~\bibnamefont{Zerwas}},
  \emph{\bibinfo{title}{{Stop Reconstruction with Tagged Tops}}}
  (\bibinfo{year}{2010}{\natexlab{a}}), \eprint{1006.2833}.

\bibitem[{\citenamefont{Ellis et~al.}(2009)\citenamefont{Ellis, Vermilion, and
  Walsh}}]{Ellis:2009su}
\bibinfo{author}{\bibfnamefont{S.~D.} \bibnamefont{Ellis}},
  \bibinfo{author}{\bibfnamefont{C.~K.} \bibnamefont{Vermilion}},
  \bibnamefont{and} \bibinfo{author}{\bibfnamefont{J.~R.} \bibnamefont{Walsh}},
  \emph{\bibinfo{title}{{Techniques for Improved Heavy Particle Searches with
  Jet Substructure}}}, \bibinfo{journal}{Phys. Rev.}
  \textbf{\bibinfo{volume}{D80}}, \bibinfo{pages}{051501}
  (\bibinfo{year}{2009}), \eprint{0903.5081}.

\bibitem[{\citenamefont{Thaler and {Van Tilburg}}(2010)}]{Thaler:2010tr}
\bibinfo{author}{\bibfnamefont{J.}~\bibnamefont{Thaler}} \bibnamefont{and}
  \bibinfo{author}{\bibfnamefont{K.}~\bibnamefont{{Van Tilburg}}},
  \emph{\bibinfo{title}{{Identifying Boosted Objects with N-subjettiness}}}
  (\bibinfo{year}{2010}), \eprint{1011.2268}.

\bibitem[{\citenamefont{Soper and Spannowsky}(2013)}]{Soper:2012pb}
\bibinfo{author}{\bibfnamefont{D.~E.} \bibnamefont{Soper}} \bibnamefont{and}
  \bibinfo{author}{\bibfnamefont{M.}~\bibnamefont{Spannowsky}},
  \emph{\bibinfo{title}{{Finding top quarks with shower deconstruction}}},
  \bibinfo{journal}{Phys.Rev.} \textbf{\bibinfo{volume}{D87}},
  \bibinfo{pages}{054012} (\bibinfo{year}{2013}), \eprint{1211.3140}.

\bibitem[{\citenamefont{Schaetzel and Spannowsky}(2013)}]{Schaetzel:2013vka}
\bibinfo{author}{\bibfnamefont{S.}~\bibnamefont{Schaetzel}} \bibnamefont{and}
  \bibinfo{author}{\bibfnamefont{M.}~\bibnamefont{Spannowsky}},
  \emph{\bibinfo{title}{{Tagging highly boosted top quarks}}}
  (\bibinfo{year}{2013}), \eprint{1308.0540}.

\bibitem[{\citenamefont{Abdesselam et~al.}(2011)\citenamefont{Abdesselam,
  Kuutmann, Bitenc, Brooijmans, Butterworth et~al.}}]{Abdesselam:2010pt}
\bibinfo{author}{\bibfnamefont{A.}~\bibnamefont{Abdesselam}},
  \bibinfo{author}{\bibfnamefont{E.~B.} \bibnamefont{Kuutmann}},
  \bibinfo{author}{\bibfnamefont{U.}~\bibnamefont{Bitenc}},
  \bibinfo{author}{\bibfnamefont{G.}~\bibnamefont{Brooijmans}},
  \bibinfo{author}{\bibfnamefont{J.}~\bibnamefont{Butterworth}},
  \bibnamefont{et~al.}, \emph{\bibinfo{title}{{Boosted objects: A Probe of
  beyond the Standard Model physics}}}, \bibinfo{journal}{Eur.Phys.J.}
  \textbf{\bibinfo{volume}{C71}}, \bibinfo{pages}{1661} (\bibinfo{year}{2011}),
  \eprint{1012.5412}.

\bibitem[{\citenamefont{Altheimer et~al.}(2012)\citenamefont{Altheimer, Arora,
  Asquith, Brooijmans, Butterworth et~al.}}]{Altheimer:2012mn}
\bibinfo{author}{\bibfnamefont{A.}~\bibnamefont{Altheimer}},
  \bibinfo{author}{\bibfnamefont{S.}~\bibnamefont{Arora}},
  \bibinfo{author}{\bibfnamefont{L.}~\bibnamefont{Asquith}},
  \bibinfo{author}{\bibfnamefont{G.}~\bibnamefont{Brooijmans}},
  \bibinfo{author}{\bibfnamefont{J.}~\bibnamefont{Butterworth}},
  \bibnamefont{et~al.}, \emph{\bibinfo{title}{{Jet Substructure at the Tevatron
  and LHC: New results, new tools, new benchmarks}}},
  \bibinfo{journal}{J.Phys.} \textbf{\bibinfo{volume}{G39}},
  \bibinfo{pages}{063001} (\bibinfo{year}{2012}), \eprint{1201.0008}.

\bibitem[{\citenamefont{Seymour}(1994)}]{Seymour:1993mx}
\bibinfo{author}{\bibfnamefont{M.~H.} \bibnamefont{Seymour}},
  \emph{\bibinfo{title}{{Searches for New Particles Using Cone and Cluster Jet
  Algorithms: A Comparative Study}}}, \bibinfo{journal}{Z. Phys.}
  \textbf{\bibinfo{volume}{C62}}, \bibinfo{pages}{127} (\bibinfo{year}{1994}).

\bibitem[{\citenamefont{Butterworth et~al.}(2002)\citenamefont{Butterworth,
  Cox, and Forshaw}}]{Butterworth:2002tt}
\bibinfo{author}{\bibfnamefont{J.~M.} \bibnamefont{Butterworth}},
  \bibinfo{author}{\bibfnamefont{B.~E.} \bibnamefont{Cox}}, \bibnamefont{and}
  \bibinfo{author}{\bibfnamefont{J.~R.} \bibnamefont{Forshaw}},
  \emph{\bibinfo{title}{{$WW$ Scattering at the CERN LHC}}},
  \bibinfo{journal}{Phys. Rev.} \textbf{\bibinfo{volume}{D65}},
  \bibinfo{pages}{096014} (\bibinfo{year}{2002}), \eprint{hep-ph/0201098}.

\bibitem[{\citenamefont{Butterworth et~al.}(2008)\citenamefont{Butterworth,
  Davison, Rubin, and Salam}}]{Butterworth:2008iy}
\bibinfo{author}{\bibfnamefont{J.~M.} \bibnamefont{Butterworth}},
  \bibinfo{author}{\bibfnamefont{A.~R.} \bibnamefont{Davison}},
  \bibinfo{author}{\bibfnamefont{M.}~\bibnamefont{Rubin}}, \bibnamefont{and}
  \bibinfo{author}{\bibfnamefont{G.~P.} \bibnamefont{Salam}},
  \emph{\bibinfo{title}{{Jet Substructure as a New Higgs Search Channel at the
  LHC}}}, \bibinfo{journal}{Phys.Rev.Lett.} \textbf{\bibinfo{volume}{100}},
  \bibinfo{pages}{242001} (\bibinfo{year}{2008}), \eprint{0802.2470}.

\bibitem[{\citenamefont{Krohn et~al.}(2010)\citenamefont{Krohn, Thaler, and
  Wang}}]{Krohn:2009th}
\bibinfo{author}{\bibfnamefont{D.}~\bibnamefont{Krohn}},
  \bibinfo{author}{\bibfnamefont{J.}~\bibnamefont{Thaler}}, \bibnamefont{and}
  \bibinfo{author}{\bibfnamefont{L.-T.} \bibnamefont{Wang}},
  \emph{\bibinfo{title}{{Jet Trimming}}}, \bibinfo{journal}{JHEP}
  \textbf{\bibinfo{volume}{02}}, \bibinfo{pages}{084} (\bibinfo{year}{2010}),
  \eprint{0912.1342}.

\bibitem[{\citenamefont{Soper and Spannowsky}(2010)}]{Soper:2010xk}
\bibinfo{author}{\bibfnamefont{D.~E.} \bibnamefont{Soper}} \bibnamefont{and}
  \bibinfo{author}{\bibfnamefont{M.}~\bibnamefont{Spannowsky}},
  \emph{\bibinfo{title}{{Combining subjet algorithms to enhance ZH detection at
  the LHC}}}, \bibinfo{journal}{JHEP} \textbf{\bibinfo{volume}{1008}},
  \bibinfo{pages}{029} (\bibinfo{year}{2010}), \eprint{1005.0417}.

\bibitem[{\citenamefont{Cui et~al.}(2011)\citenamefont{Cui, Han, and
  Schwartz}}]{Cui:2010km}
\bibinfo{author}{\bibfnamefont{Y.}~\bibnamefont{Cui}},
  \bibinfo{author}{\bibfnamefont{Z.}~\bibnamefont{Han}}, \bibnamefont{and}
  \bibinfo{author}{\bibfnamefont{M.~D.} \bibnamefont{Schwartz}},
  \emph{\bibinfo{title}{{W-jet Tagging: Optimizing the Identification of
  Boosted Hadronically-Decaying W Bosons}}}, \bibinfo{journal}{Phys.Rev.}
  \textbf{\bibinfo{volume}{D83}}, \bibinfo{pages}{074023}
  (\bibinfo{year}{2011}), \eprint{1012.2077}.

\bibitem[{\citenamefont{Hackstein and Spannowsky}(2010)}]{Hackstein:2010wk}
\bibinfo{author}{\bibfnamefont{C.}~\bibnamefont{Hackstein}} \bibnamefont{and}
  \bibinfo{author}{\bibfnamefont{M.}~\bibnamefont{Spannowsky}},
  \emph{\bibinfo{title}{{Boosting Higgs Discovery - The Forgotten Channel}}}
  (\bibinfo{year}{2010}), \eprint{1008.2202}.

\bibitem[{\citenamefont{Soper and Spannowsky}(2011)}]{Soper:2011cr}
\bibinfo{author}{\bibfnamefont{D.~E.} \bibnamefont{Soper}} \bibnamefont{and}
  \bibinfo{author}{\bibfnamefont{M.}~\bibnamefont{Spannowsky}},
  \emph{\bibinfo{title}{{Finding physics signals with shower deconstruction}}},
  \bibinfo{journal}{Phys.Rev.} \textbf{\bibinfo{volume}{D84}},
  \bibinfo{pages}{074002} (\bibinfo{year}{2011}), \eprint{1102.3480}.

\bibitem[{\citenamefont{Almeida et~al.}(2012)\citenamefont{Almeida, Erdogan,
  Juknevich, Lee, Perez et~al.}}]{Almeida:2011aa}
\bibinfo{author}{\bibfnamefont{L.~G.} \bibnamefont{Almeida}},
  \bibinfo{author}{\bibfnamefont{O.}~\bibnamefont{Erdogan}},
  \bibinfo{author}{\bibfnamefont{J.}~\bibnamefont{Juknevich}},
  \bibinfo{author}{\bibfnamefont{S.~J.} \bibnamefont{Lee}},
  \bibinfo{author}{\bibfnamefont{G.}~\bibnamefont{Perez}},
  \bibnamefont{et~al.}, \emph{\bibinfo{title}{{Three-particle templates for a
  boosted Higgs boson}}}, \bibinfo{journal}{Phys.Rev.}
  \textbf{\bibinfo{volume}{D85}}, \bibinfo{pages}{114046}
  (\bibinfo{year}{2012}), \eprint{1112.1957}.

\bibitem[{\citenamefont{Dasgupta et~al.}(2013)\citenamefont{Dasgupta, Fregoso,
  Marzani, and Salam}}]{Dasgupta:2013ihk}
\bibinfo{author}{\bibfnamefont{M.}~\bibnamefont{Dasgupta}},
  \bibinfo{author}{\bibfnamefont{A.}~\bibnamefont{Fregoso}},
  \bibinfo{author}{\bibfnamefont{S.}~\bibnamefont{Marzani}}, \bibnamefont{and}
  \bibinfo{author}{\bibfnamefont{G.~P.} \bibnamefont{Salam}},
  \emph{\bibinfo{title}{{Towards an understanding of jet substructure}}}
  (\bibinfo{year}{2013}), \eprint{1307.0007}.

\bibitem[{\citenamefont{Harigaya et~al.}(2012)\citenamefont{Harigaya,
  Matsumoto, Nojiri, and Tobioka}}]{Harigaya:2012ir}
\bibinfo{author}{\bibfnamefont{K.}~\bibnamefont{Harigaya}},
  \bibinfo{author}{\bibfnamefont{S.}~\bibnamefont{Matsumoto}},
  \bibinfo{author}{\bibfnamefont{M.~M.} \bibnamefont{Nojiri}},
  \bibnamefont{and} \bibinfo{author}{\bibfnamefont{K.}~\bibnamefont{Tobioka}},
  \emph{\bibinfo{title}{{Search for the Top Partner at the LHC using
  Multi-b-Jet Channels}}}, \bibinfo{journal}{Phys.Rev.}
  \textbf{\bibinfo{volume}{D86}}, \bibinfo{pages}{015005}
  (\bibinfo{year}{2012}), \eprint{1204.2317}.

\bibitem[{\citenamefont{Contino
  et~al.}(2007{\natexlab{b}})\citenamefont{Contino, Kramer, Son, and
  Sundrum}}]{Contino:2006nn}
\bibinfo{author}{\bibfnamefont{R.}~\bibnamefont{Contino}},
  \bibinfo{author}{\bibfnamefont{T.}~\bibnamefont{Kramer}},
  \bibinfo{author}{\bibfnamefont{M.}~\bibnamefont{Son}}, \bibnamefont{and}
  \bibinfo{author}{\bibfnamefont{R.}~\bibnamefont{Sundrum}},
  \emph{\bibinfo{title}{{Warped/Composite Phenomenology Simplified}}},
  \bibinfo{journal}{JHEP} \textbf{\bibinfo{volume}{05}}, \bibinfo{pages}{074}
  (\bibinfo{year}{2007}{\natexlab{b}}), \eprint{hep-ph/0612180}.

\bibitem[{\citenamefont{Contino and Servant}(2008)}]{Contino:2008hi}
\bibinfo{author}{\bibfnamefont{R.}~\bibnamefont{Contino}} \bibnamefont{and}
  \bibinfo{author}{\bibfnamefont{G.}~\bibnamefont{Servant}},
  \emph{\bibinfo{title}{{Discovering the top partners at the LHC using
  same-sign dilepton final states}}}, \bibinfo{journal}{JHEP}
  \textbf{\bibinfo{volume}{0806}}, \bibinfo{pages}{026} (\bibinfo{year}{2008}),
  \eprint{0801.1679}.

\bibitem[{\citenamefont{Coleman et~al.}(1969)\citenamefont{Coleman, Wess, and
  Zumino}}]{Coleman:1969sm}
\bibinfo{author}{\bibfnamefont{S.~R.} \bibnamefont{Coleman}},
  \bibinfo{author}{\bibfnamefont{J.}~\bibnamefont{Wess}}, \bibnamefont{and}
  \bibinfo{author}{\bibfnamefont{B.}~\bibnamefont{Zumino}},
  \emph{\bibinfo{title}{{Structure of phenomenological Lagrangians. 1.}}},
  \bibinfo{journal}{Phys.Rev.} \textbf{\bibinfo{volume}{177}},
  \bibinfo{pages}{2239} (\bibinfo{year}{1969}).

\bibitem[{\citenamefont{Callan et~al.}(1969)\citenamefont{Callan, Coleman,
  Wess, and Zumino}}]{Callan:1969sn}
\bibinfo{author}{\bibfnamefont{J.}~\bibnamefont{Callan},
  \bibfnamefont{Curtis~G.}}, \bibinfo{author}{\bibfnamefont{S.~R.}
  \bibnamefont{Coleman}},
  \bibinfo{author}{\bibfnamefont{J.}~\bibnamefont{Wess}}, \bibnamefont{and}
  \bibinfo{author}{\bibfnamefont{B.}~\bibnamefont{Zumino}},
  \emph{\bibinfo{title}{{Structure of phenomenological Lagrangians. 2.}}},
  \bibinfo{journal}{Phys.Rev.} \textbf{\bibinfo{volume}{177}},
  \bibinfo{pages}{2247} (\bibinfo{year}{1969}).

\bibitem[{\citenamefont{Panico et~al.}(2013)\citenamefont{Panico, Redi, Tesi,
  and Wulzer}}]{Panico:2012uw}
\bibinfo{author}{\bibfnamefont{G.}~\bibnamefont{Panico}},
  \bibinfo{author}{\bibfnamefont{M.}~\bibnamefont{Redi}},
  \bibinfo{author}{\bibfnamefont{A.}~\bibnamefont{Tesi}}, \bibnamefont{and}
  \bibinfo{author}{\bibfnamefont{A.}~\bibnamefont{Wulzer}},
  \emph{\bibinfo{title}{{On the Tuning and the Mass of the Composite Higgs}}},
  \bibinfo{journal}{JHEP} \textbf{\bibinfo{volume}{1303}}, \bibinfo{pages}{051}
  (\bibinfo{year}{2013}), \eprint{1210.7114}.

\bibitem[{\citenamefont{Cacciapaglia et~al.}(2013)\citenamefont{Cacciapaglia,
  Deandrea, Panizzi, Perries, and Sordini}}]{Cacciapaglia:2012dd}
\bibinfo{author}{\bibfnamefont{G.}~\bibnamefont{Cacciapaglia}},
  \bibinfo{author}{\bibfnamefont{A.}~\bibnamefont{Deandrea}},
  \bibinfo{author}{\bibfnamefont{L.}~\bibnamefont{Panizzi}},
  \bibinfo{author}{\bibfnamefont{S.}~\bibnamefont{Perries}}, \bibnamefont{and}
  \bibinfo{author}{\bibfnamefont{V.}~\bibnamefont{Sordini}},
  \emph{\bibinfo{title}{{Heavy Vector-like quark with charge 5/3 at the LHC}}},
  \bibinfo{journal}{JHEP} \textbf{\bibinfo{volume}{1303}}, \bibinfo{pages}{004}
  (\bibinfo{year}{2013}), \eprint{1211.4034}.

\bibitem[{\citenamefont{Degrande et~al.}(2012)\citenamefont{Degrande, Duhr,
  Fuks, Grellscheid, Mattelaer et~al.}}]{Degrande:2011ua}
\bibinfo{author}{\bibfnamefont{C.}~\bibnamefont{Degrande}},
  \bibinfo{author}{\bibfnamefont{C.}~\bibnamefont{Duhr}},
  \bibinfo{author}{\bibfnamefont{B.}~\bibnamefont{Fuks}},
  \bibinfo{author}{\bibfnamefont{D.}~\bibnamefont{Grellscheid}},
  \bibinfo{author}{\bibfnamefont{O.}~\bibnamefont{Mattelaer}},
  \bibnamefont{et~al.}, \emph{\bibinfo{title}{{UFO - The Universal FeynRules
  Output}}}, \bibinfo{journal}{Comput.Phys.Commun.}
  \textbf{\bibinfo{volume}{183}}, \bibinfo{pages}{1201} (\bibinfo{year}{2012}),
  \eprint{1108.2040}.

\bibitem[{\citenamefont{Cacciari and Salam}(2006)}]{Cacciari:2005hq}
\bibinfo{author}{\bibfnamefont{M.}~\bibnamefont{Cacciari}} \bibnamefont{and}
  \bibinfo{author}{\bibfnamefont{G.~P.} \bibnamefont{Salam}},
  \emph{\bibinfo{title}{{Dispelling the $N^{3}$ Myth for the $k\_t$
  Jet-Finder}}}, \bibinfo{journal}{Phys. Lett.}
  \textbf{\bibinfo{volume}{B641}}, \bibinfo{pages}{57} (\bibinfo{year}{2006}),
  \eprint{hep-ph/0512210}.

\bibitem[{\citenamefont{Aliev et~al.}(2011)\citenamefont{Aliev, Lacker,
  Langenfeld, Moch, Uwer et~al.}}]{Aliev:2010zk}
\bibinfo{author}{\bibfnamefont{M.}~\bibnamefont{Aliev}},
  \bibinfo{author}{\bibfnamefont{H.}~\bibnamefont{Lacker}},
  \bibinfo{author}{\bibfnamefont{U.}~\bibnamefont{Langenfeld}},
  \bibinfo{author}{\bibfnamefont{S.}~\bibnamefont{Moch}},
  \bibinfo{author}{\bibfnamefont{P.}~\bibnamefont{Uwer}}, \bibnamefont{et~al.},
  \emph{\bibinfo{title}{{HATHOR: HAdronic Top and Heavy quarks crOss section
  calculatoR}}}, \bibinfo{journal}{Comput.Phys.Commun.}
  \textbf{\bibinfo{volume}{182}}, \bibinfo{pages}{1034} (\bibinfo{year}{2011}),
  \eprint{1007.1327}.

\bibitem[{\citenamefont{Alwall et~al.}(2008)\citenamefont{Alwall, Hoche,
  Krauss, Lavesson, Lonnblad et~al.}}]{Alwall:2007fs}
\bibinfo{author}{\bibfnamefont{J.}~\bibnamefont{Alwall}},
  \bibinfo{author}{\bibfnamefont{S.}~\bibnamefont{Hoche}},
  \bibinfo{author}{\bibfnamefont{F.}~\bibnamefont{Krauss}},
  \bibinfo{author}{\bibfnamefont{N.}~\bibnamefont{Lavesson}},
  \bibinfo{author}{\bibfnamefont{L.}~\bibnamefont{Lonnblad}},
  \bibnamefont{et~al.}, \emph{\bibinfo{title}{{Comparative study of various
  algorithms for the merging of parton showers and matrix elements in hadronic
  collisions}}}, \bibinfo{journal}{Eur.Phys.J.} \textbf{\bibinfo{volume}{C53}},
  \bibinfo{pages}{473} (\bibinfo{year}{2008}), \eprint{0706.2569}.

\bibitem[{\citenamefont{Dokshitzer et~al.}(1997)\citenamefont{Dokshitzer,
  Leder, Moretti, and Webber}}]{Dokshitzer:1997in}
\bibinfo{author}{\bibfnamefont{Y.~L.} \bibnamefont{Dokshitzer}},
  \bibinfo{author}{\bibfnamefont{G.~D.} \bibnamefont{Leder}},
  \bibinfo{author}{\bibfnamefont{S.}~\bibnamefont{Moretti}}, \bibnamefont{and}
  \bibinfo{author}{\bibfnamefont{B.~R.} \bibnamefont{Webber}},
  \emph{\bibinfo{title}{{Better Jet Clustering Algorithms}}},
  \bibinfo{journal}{JHEP} \textbf{\bibinfo{volume}{08}}, \bibinfo{pages}{001}
  (\bibinfo{year}{1997}), \eprint{hep-ph/9707323}.

\bibitem[{\citenamefont{Wobisch and Wengler}(1998)}]{Wobisch:1998wt}
\bibinfo{author}{\bibfnamefont{M.}~\bibnamefont{Wobisch}} \bibnamefont{and}
  \bibinfo{author}{\bibfnamefont{T.}~\bibnamefont{Wengler}},
  \emph{\bibinfo{title}{{Hadronization Corrections to Jet Cross Sections in
  Deep- Inelastic Scattering}}} (\bibinfo{year}{1998}),
  \eprint{hep-ph/9907280}.

\bibitem[{\citenamefont{Plehn et~al.}(2010{\natexlab{b}})\citenamefont{Plehn,
  Salam, and Spannowsky}}]{Plehn:2009rk}
\bibinfo{author}{\bibfnamefont{T.}~\bibnamefont{Plehn}},
  \bibinfo{author}{\bibfnamefont{G.~P.} \bibnamefont{Salam}}, \bibnamefont{and}
  \bibinfo{author}{\bibfnamefont{M.}~\bibnamefont{Spannowsky}},
  \emph{\bibinfo{title}{{Fat Jets for a Light Higgs}}}, \bibinfo{journal}{Phys.
  Rev. Lett.} \textbf{\bibinfo{volume}{104}}, \bibinfo{pages}{111801}
  (\bibinfo{year}{2010}{\natexlab{b}}), \eprint{0910.5472}.

\bibitem[{\citenamefont{Cacciari et~al.}(2008)\citenamefont{Cacciari, Salam,
  and Soyez}}]{Cacciari:2008gp}
\bibinfo{author}{\bibfnamefont{M.}~\bibnamefont{Cacciari}},
  \bibinfo{author}{\bibfnamefont{G.~P.} \bibnamefont{Salam}}, \bibnamefont{and}
  \bibinfo{author}{\bibfnamefont{G.}~\bibnamefont{Soyez}},
  \emph{\bibinfo{title}{{The Anti-$k_T$ Jet Clustering Algorithm}}},
  \bibinfo{journal}{JHEP} \textbf{\bibinfo{volume}{04}}, \bibinfo{pages}{063}
  (\bibinfo{year}{2008}), \eprint{0802.1189}.

\bibitem[{\citenamefont{Grojean et~al.}(2013)\citenamefont{Grojean,
  Matsedonskyi, and Panico}}]{Grojean:2013qca}
\bibinfo{author}{\bibfnamefont{C.}~\bibnamefont{Grojean}},
  \bibinfo{author}{\bibfnamefont{O.}~\bibnamefont{Matsedonskyi}},
  \bibnamefont{and} \bibinfo{author}{\bibfnamefont{G.}~\bibnamefont{Panico}},
  \emph{\bibinfo{title}{{Light top partners and precision physics}}}
  (\bibinfo{year}{2013}), \eprint{1306.4655}.

\bibitem[{\citenamefont{Rehermann and Tweedie}(2011)}]{Rehermann:2010vq}
\bibinfo{author}{\bibfnamefont{K.}~\bibnamefont{Rehermann}} \bibnamefont{and}
  \bibinfo{author}{\bibfnamefont{B.}~\bibnamefont{Tweedie}},
  \emph{\bibinfo{title}{{Efficient Identification of Boosted Semileptonic Top
  Quarks at the LHC}}}, \bibinfo{journal}{JHEP}
  \textbf{\bibinfo{volume}{1103}}, \bibinfo{pages}{059} (\bibinfo{year}{2011}),
  \eprint{1007.2221}.

\end{thebibliography}
\bibliographystyle{apsper}

\end{document}